\documentclass[conference]{IEEEtran} 

\usepackage[T1]{fontenc} 
\usepackage{amsmath,amssymb,physics} 
\usepackage{algorithm}
\usepackage{algpseudocode}
\usepackage{float}
\usepackage{graphicx} 
\usepackage{cite} 
\usepackage{balance} 
\usepackage{qcircuit}
\usepackage{subcaption}
\usepackage{todonotes}
\usepackage{comment}
\usepackage{siunitx}
\usepackage{booktabs}
\usepackage{blindtext}
\usepackage{svg}
\usepackage{url}
\usepackage{xurl}

\usepackage{hyperref,xcolor}
\definecolor{ionqorange}{HTML}{FF5000}
\hypersetup{
    colorlinks=true,
    linkcolor=ionqorange,    
    citecolor=ionqorange,    
    urlcolor=ionqorange      
}

\IEEEoverridecommandlockouts 

\title{Accelerating Large-Scale Linear Algebra  Using Variational Quantum Imaginary Time Evolution}
\author{
    \IEEEauthorblockN{
        Willie Aboumrad\IEEEauthorrefmark{1}\textsuperscript{\textsection}, 
        Daiwei Zhu\IEEEauthorrefmark{1}\textsuperscript{\textsection}, 
        Claudio Girotto\IEEEauthorrefmark{1},
        Fran\c cois-Henry Rouet\IEEEauthorrefmark{2},
        Jezer Jojo\IEEEauthorrefmark{2},
    }
    \IEEEauthorblockN{
        Robert Lucas\IEEEauthorrefmark{2},
        Jay Pathak\IEEEauthorrefmark{2},
        Ananth Kaushik\IEEEauthorrefmark{1},
        Martin Roetteler\IEEEauthorrefmark{1}
    }
    \IEEEauthorblockA{\IEEEauthorrefmark{1}IonQ Inc., 4505 Campus Dr, College Park, MD 20740, USA}
    \IEEEauthorblockA{\IEEEauthorrefmark{2}Ansys, Inc., 2600 Ansys Drive, Canonsburg, PA 15317, USA}
}

\begin{document}

\maketitle
\begingroup\renewcommand\thefootnote{\textsection}
\footnotetext{These authors contributed equally to this work}
\endgroup

\begin{abstract}

    The solution of large sparse linear systems via factorization methods such as LU or Cholesky decomposition, can be computationally expensive due to the introduction of non-zero elements, or ``fill-in.'' Graph partitioning can be used to reduce the ``fill-in,'' thereby speeding up the solution of the linear system. We introduce a quantum approach to the graph partitioning problem based on variational quantum imaginary time evolution (VarQITE). We develop a hybrid quantum/classical method to accelerate Finite Element Analysis (FEA) by using VarQITE in Ansys's LS-DYNA multiphysics simulation software. This allows us to study different types of FEA problems, from mechanical engineering to computational fluid dynamics in simulations and on quantum hardware (IonQ Aria and IonQ Forte). 
    
    We demonstrate that VarQITE has the potential to impact LS-DYNA workflows by measuring the wall-clock time to solution of FEA problems. We report performance results for our hybrid quantum/classical workflow on selected FEA problem instances, including simulation of blood pumps, automotive roof crush, and vibration analysis of car bodies on meshes of up to six million elements. We find that the LS-DYNA wall clock time can be improved by up to 12\% for some problems. Finally, we introduce a classical heuristic inspired by Fiduccia-Mattheyses to improve the quality of VarQITE solutions obtained from hardware runs. 
    Our results highlight the potential impact of quantum computing on large-scale FEA problems in the NISQ era.\\[1ex]
\end{abstract}

\section{Introduction}

{Solving a linear system $Ax = b$, with $A$ and $b$ known, is the main computational bottleneck in a number of scientific and commercial computing applications. The matrix $A$ is typically large and \emph{sparse}, with representative cases having  tens to hundreds of millions of rows and columns, yet only a small number of entries per row are non-zero. Such matrices are typically distributed across multiple compute nodes, and only the non-zero entries are stored. Mechanical simulations typically use \emph{direct} solvers, which solve the problem by factoring $A$ into upper and lower triangular matrices, allowing the linear system to be easily solved with forward elimination and backward substitution.} 

{The factorization step eliminates non-zero entries in the lower triangle of the matrix by subtracting previous rows.  Zeroing out one entry often introduces new non-zero entries, a process called fill-in. Fill-in dynamically increases the computational burden of the factorization step. In practice, finding an optimal reordering of a sparse matrix that minimizes fill-in is a combinatorial optimization problem, i.e., an NP-Complete problem \cite{yannakakis1981computing}.  Finding permutations to reduce fill-in is especially difficult when matrices are distributed over thousands of processors. As the processor count increases, the reordering step can dominate the entire run-time of an FEA application.}

{The reordering problem is well-studied and many classical heuristics are available. In computational mechanics, ``Nested Dissection,'' a divide-and-conquer approach introduced in 1973 \cite{george1973nested}, remains the most successful. In this method, the sparse matrix is represented as an undirected graph in which the vertices represent rows and columns of the matrix and the edges represent the non-zero entries. This graph is recursively partitioned into subgraphs using \emph{separators}, small subsets of vertices whose removal allows the graph to be partitioned into disjoint subgraphs with at most a constant fraction of the number of vertices. Most implementations of Nested Dissection, e.g., METIS~\cite{METIS}, rely on a so-called \emph{multilevel} scheme, where the graph to be partitioned is first \emph{coarsened} into a smaller graph. The coarse graph is partitioned into two parts, and the partition is then \emph{projected} onto the original graph and transformed into a vertex separator before being refined. The bipartition of the coarse graph is a critical component and the focus of our study. }

{The Graph Partitioning Problem (GPP) has an long history in theoretical computer science, and countless applications in scientific computing, task scheduling, social networks, VLSI design, etc. The GPP seeks to partition the vertices of a graph into balanced parts, while minimizing an objective function such as the number of cut edges. It is known that no polynomial time algorithm can obtain balanced graph partitions within a finite approximation factor~\cite{andreev2006gpp}. A range of heuristic algorithms have been developed that can efficiently produce solutions in practical settings. In this work we explore the potential of quantum algorithms to obtain higher quality graph partitions, exploiting quantum resources to improve upon leading classical heuristics. In particular, we adapt a variational quantum imaginary time evolution (VarQITE) algorithm \cite{Morris2024-gj} for approximately solving multiple GPPs constructed by Ansys's LS-DYNA~\cite{Ansys_LSDYNA} commercial multiphysics simulation software. 

Concretely, we demonstrate the applicability of our novel hybrid quantum-classical algorithm by integrating quantum computations executed on IonQ simulators as well as IonQ's Aria and Forte quantum hardare into a large-scale simulations conducted by LS-DYNA, and discuss encouraging results highlighting the potential for quantum utility in the NISQ era.}

The rest of this paper is organized as follows. In Section~\ref{sec:background}, we describe the GPP in some detail, along with its formulation as a Hamiltonian Energy Minimization problem. In addition, we briefly describe the VarQITE algorithm. In Section~\ref{sec:methods} we explain the structure and formulation of our quantum ansatz as well as how to apply VarQITE to obtain high-quality approximations of the GPP. We also describe the pipeline of integration of VarQITE into LS-DYNA and the metrics for evaluation and comparison of solutions to the classical heuristic graph partitioning solver in LS-DYNA (LS-GPart). In Section~\ref{sec:results} we discuss the results of our quantum computations both from noiseless simulations as well as from executions on IonQ quantum hardware Aria and Forte. Finally, in Section~\ref{sec:future-outlook} we discuss ongoing research and lay out an optimistic outlook for the future development of our quantum-classical methods.

\section{Problem formulation}
\label{sec:background}

\subsection{The Graph Partitioning Problem (GPP) as a Quadratic Program (QP)}

Given a (vertex- and edge-) weighted graph $G = (V, E)$, a \(k\)-way partition of the graph is a family of disjoint subsets \(V_1,V_2,\ldots,V_k\) of \(V\) s.t. \(V_1\sqcup V_2\ldots\sqcup V_k = V\). When \(k=2\), this is referred to as a \emph{bipartition}. Different Graph Partitioning Problems (GPP) can be formulated based on the objective function to minimize and constraints placed on the partition. A typical objective function is to minimize the sum of the weight of \emph{cut edges}, edges which connect vertices belonging to different parts, i.\,e., to the set $\{((v_i, v_j)\in E\text{ with }v_i\in V_i, v_j\in V_j, i\neq j\}$. A typical constraint is to enforce that parts have equal or similar sizes. In the context of multilevel Nested Dissection, there is freedom in the choice of the constraints. Our choice is to find a bipartition \(\{V_1, V_2\}\) such that \(|V_1|\) and \(|V_2|\) do not exceed $({1 \over 2} + \nu)|V|$ where \(|\cdot|\) denotes the sum of vertex weights. In this paper, we set $\nu = 0.05$.

In symbols, we formulate our GPP as the following QP:

\begin{align}\label{eqn:gpp-defn}
    \mathrm{minimize}_x &\sum_{\substack{(i,j) \in E}} w_{ij}(x_i + x_j - 2x_i x_j), \nonumber \\
    \mathrm{s.t.} & \sum_i v_i x_i \leq \big(1/2 + \nu\big) \Omega \\
                  & \sum_i v_i (1 - x_i) \leq \big(1/2 + \nu\big) \Omega \nonumber\\
                  & x_i \in \{0, 1\}. \nonumber
\end{align}

Here, the $w_{ij}$ denote the edge weights, the $v_i$ denote the node weights, $\Omega$ denotes the total sum of the node weights, and the Boolean variable $x_i$ determines whether the $i$th node belongs to $V_1$ ($x_i=0$) or $V_2$ ($x_i=1$). 

\subsection{GPP as a Quadratic Unconstrained Binary Optimization (QUBO) problem}

In order to leverage quantum resources for solving our GPP, we first formulate the QP in equation~(\ref{eqn:gpp-defn}) as a QUBO. Standard techniques convert inequality constraints into penalty terms by introducing slack variables. This translates into higher qubit requirements when minimizing on quantum hardware. We opt for the following penalty instead:

\begin{equation}
    P(x) = \left(\sum_{i} v_ix_i - \sum_{i} \frac{v_i}{2}\right)^2.
\end{equation}

This quadratic term penalizes deviations in the total node weight on each side of the partition; it is minimized when the sum of the node weights in the two partitions are equal. 

Thus, the problem at hand is to minimize the unconstrained objective

\begin{align}
    C(x) = \sum_{\substack{(i,j) \in E}} w_{ij}(x_i + x_j - 2x_i x_j) + \lambda P(x).
\end{align}

Here $\lambda>0$ is a tunable hyper-parameter that defines a trade-off between lower weight and more balanced cuts.

\subsection{From GPP to Hamiltonian Energy Minimization}\label{sec:HQUBO}

Given $C(x)$, we encode our GPP as a Hamiltonian Energy Minimization problem by constructing a Hamiltonian $H_C$ on $n = |V|$ qubits such that
$$
    H_C \ket{x} = C(x) \ket{x},
$$
using standard techniques. In particular, we obtain $H_C$ by replacing each $x_j$ in the expression of $C(x)$ by the operator
$$
    \hat{N}_j = \frac{1}{2}(I - Z_j),
$$
where $I$ denotes the identity operator on $n$ qubits and $Z_j$ denotes the
Pauli-Z operator acting on the $j$th qubit. Notice that $\hat{N}_j$ is
diagonal with respect to the computational basis, and its eigenvalues are zero
and one; in particular,
$$
    \hat{N}_j \ket{x} = x_j \ket{x}.
$$

\subsection{Variational Quantum Imaginary Time Evolution}
\label{varQITE}

Quantum imaginary time evolution (QITE) is a powerful approach to compute the ground state of a physical system. It expresses the ground-state as the long-time limit of the imaginary time Schr\"{o}dinger equation. Imaginary time evolution, a concept originating in quantum many-body physics, systematically projects a given initial state $|\Psi(0)\rangle$ onto the ground state of $H_C$ by evolving it in imaginary time $\tau = it$, where $t$ is the real time parameter. This process is governed by the imaginary-time Schrödinger equation
\begin{align}
    \pdv{\tau}\ket{\Psi(\tau)} = -(H_C - E_\tau) \ket{\Psi(\tau)},
\end{align}
where $E_\tau = \ev{H}{\Psi(\tau)}$ is the instantaneous energy resulting from enforcing normalization. For a time-independent Hamiltonian $H_C$, the solution is given by
\begin{align}
    \ket{\Psi(\tau)} = \frac{e^{-\tau H_C}}{\sqrt{\ev{e^{-2\tau H_C}}{\Psi(0)}}}\ket{\Psi(0)}. 
\end{align}
As $\tau \to \infty$, and assuming that $\ket{\Psi(0)}$ has a non-zero overlap with the ground state $\ket{\Psi_{\text{GS}}}$, this evolution exponentially suppresses excited-state components, thus converging to $\ket{\Psi_\text{GS}}$.

Since directly implementing the operator $\exp(-H_C \tau)$ on a quantum computer is challenging, we can instead approximate imaginary time evolution variationally. We introduce a parameterized ansatz $\ket{\Psi(\theta)}$ with parameters that change as a function of iteration number. The goal is to adjust these parameters $\theta(\tau)$ such that the state $\ket{\Psi(\theta(\tau))}$ closely approximates the true imaginary-time evolved state $\ket{\Psi(\tau)}$.
We adopt the approach presented in \cite{Morris2024-gj}, which enforces the imaginary time evolution of each Pauli term $P_\alpha$ in $H_C$ by
\begin{align}
\nonumber
    \pdv{\expval{P_\alpha}}{\tau} &= \sum_j 2\Re\left[ \bra{\Psi(\theta)} P_{\alpha}\, \pdv{\theta_j}\ket{\Psi(\theta)}\right] \dot{\theta}_j\\
    &= -\mel{ \Psi(\theta)}{\{P_{\alpha},H_c-E_{\tau}\}}{\Psi(\theta)},
\end{align}
with $\dot{\theta}_j = \partial \theta_j / \partial \tau$. Applying this condition to all $P_\alpha$ terms in the Hamiltonian $H_C$ yields the linear system 
\begin{align}
    G \cdot \dot{\theta} = D,
    \label{eq:varqite_soe}
\end{align}
where
\begin{align}
    G_{\alpha,j} &= \Re\left[ \bra{\Psi(\theta)} P_{\alpha}\, \pdv{\theta_j}\ket{\Psi(\theta)}\right], \\
    D_{\alpha} &= -\frac{1}{2}\mel{ \Psi(\theta)}{\{P_{\alpha},H_C-E_{\tau}\}}{\Psi(\theta)}.
\end{align}

Using the parameter-shift rule \cite{Mitarai2018-mu,Schuld2019-cq}, each column of $G$ can be determined from just two circuit evaluations. Moreover, if the Hamiltonian $H_C$ consists solely of Pauli-$Z$ operators (as those in QUBO formulations), all $P_\alpha$ terms commute, allowing the calculation of $D$ from a single circuit evaluation. Thus, at each time step, the parameters $\theta$ can be updated using a forward Euler method, $\theta \to \theta + \Delta\tau \dot{\theta}$, where $\dot{\theta}$ is obtained by inverting equation~(\ref{eq:varqite_soe}). This procedure requires only $2m + 1$ circuit evaluations per time step where $m$ is the number of parameters $\theta$, offering a substantial improvement in efficiency compared to previous VarQITE approaches \cite{Yuan2019-oj,McArdle2019-zp}.

\begin{figure}[hbt]
    \centering
    \includegraphics[width=1.0\linewidth]{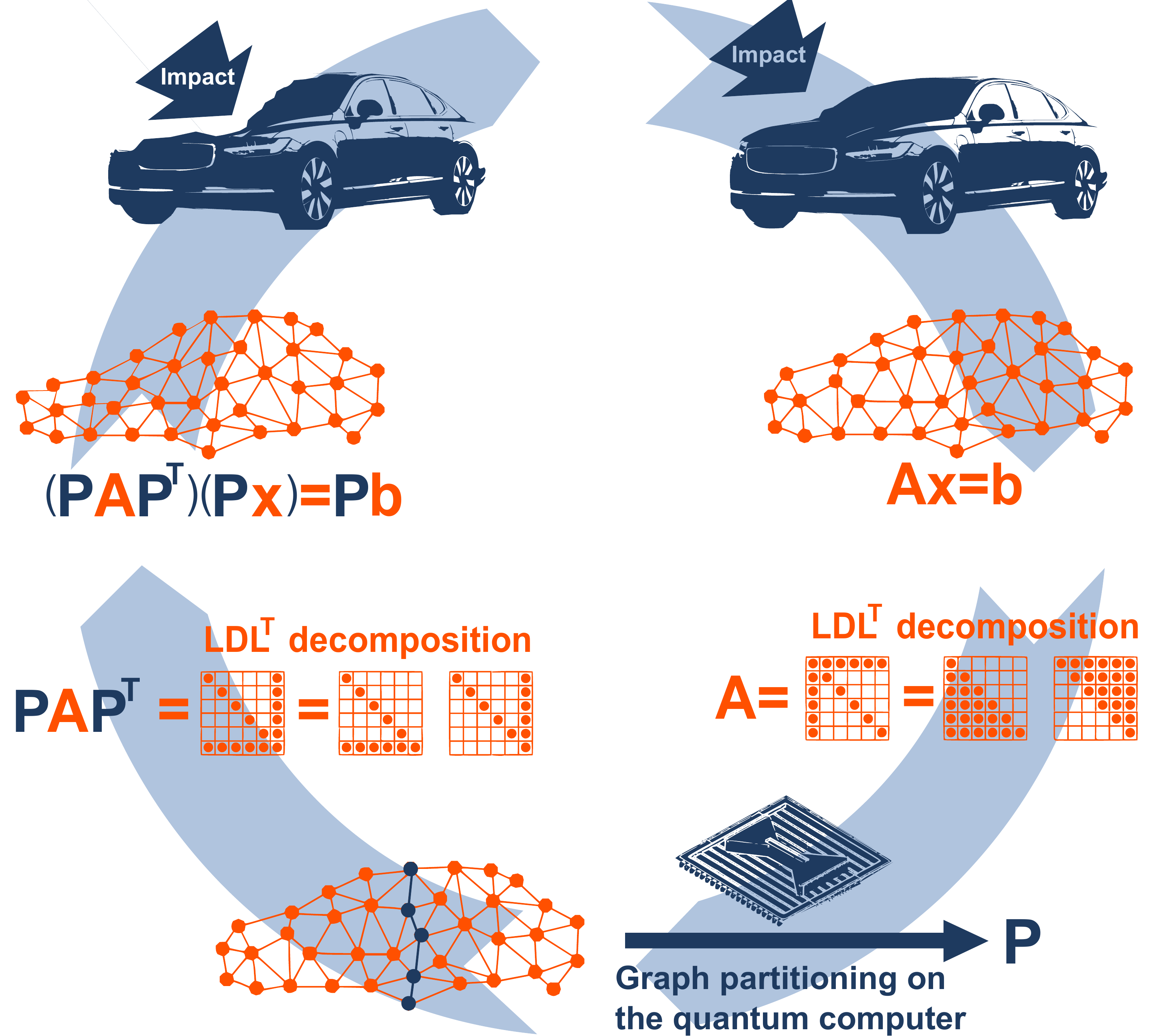}
    \caption{The execution flow chart of LS-DYNA including calls to the quantum computer. Starting from the upper right is a car which is modeled with a finite element mesh, together with an impact. The time evolution is given as a linear system of equations to be solved $Ax = b$. Here \(A\) is a sparse matrix which when decomposed using \(LDL^T\) decomposition leads to the dense matrices shown in the figure. A graph partitioning problem is formulated from the adjacency graph of \(A\) and is solved on the quantum computer, indicated by an ion trap on the lower right. This yields \(P\), a permutation matrix to reorder \(A\) obtained by solving the graph partitioning problem recursively. The solution is used to partition the mesh as illustrated by the blue line of vertices. The reordered matrix $A$ retains a sparse structure after \(LDL^T\) decomposition. The linear system is now solved using the the reordered matrix $A$ and the reordered vector $b$ to obtain the required deformation simulation of the original model. The process can be iterated. }
    \label{fig:LS-DYNA_flow_chart}
\end{figure}

\section{Methods}
\label{sec:methods}

\subsection{Application of VarQITE to the Graph Partitioning Problem}\label{subsec:problem_instances}

VarQITE can be used to solve the GPP arising in a variety of engineering FEA applications. We focus on three specific problem instances of practical interest, detailed in Table~\ref{tab:problem_instances}. Each problem has graphs with millions of edges and needs to be coarsened down to manageable sizes to be analyzed by the quantum algorithm. The LS-DYNA workflow developed by Ansys is used to perform the graph coarsening and reconstruction after the graph partitioning, as visualized in Figure~\ref{fig:LS-DYNA_flow_chart}.

\begin{table}[hbt]\setlength{\tabcolsep}{3pt}%
\centering
\resizebox{1.0\columnwidth}{!}{%
\begin{tabular}{lrrr}
\toprule
\multicolumn{1}{c}{} & RoofCrush  & BloodPump  & VibrationAnalysis  \\ 
\midrule
Vertices  &                         0.2M &                                                         0.6M &                              5.9M \\
Edges     &                         3.5M &                                                           5M &                               55M \\
Physics   &                   Structural &                                                          CFD &                        Structural \\
Mesh type &                   Tets, 2.5D &                                                 Shells, 2.5D &                        Tets, 2.5D \\
Origin    & FDA\cite{del2020performance} &  NCAC \cite{marzougui2014development,reichert2016validation} & NHTSA \& Arup\cite{singh2016update} \\
\bottomrule
\end{tabular}%
}
\caption{Summary of the problem instances analyzed in this paper. Vertices and edges are expressed in millions (M). The roof crush simulation was performed on a publicly available FEA mesh of a Toyota Yaris, the vibration analysis was performed on a Honda Accord model.}
\label{tab:problem_instances}
\end{table}

Nested dissection uses vertex separators to split the graph into smaller pieces and recurses on these smaller pieces. The total number of recursive levels of nested dissection can be set as a parameter in the LS-DYNA control card. Vertex separators are found by first performing a graph coarsening to reduce the size of the graph. The degree of coarsening required (number of vertices of the coarse graph) may also be set as a parameter for the tool; in our experiments, it is set to a few 10s of vertices so that the problem can fit on current quantum computers. The size of the coarse graph will increase with scaling the quantum hardware to larger numbers of qubits. The coarse graph has vertices and edges with large vertex and edge weights which correspond to the sum of degrees of original vertices mapped to the same coarse vertex. The graph partitioning problem is now formulated to find the best bi-partition of the graph with the least edge cut cost while maintaining nodal weight balance between the two subgraphs (within 5\% nodal weight balance). This graph partitioning optimization problem is performed on the quantum simulator/QPU hardware using VarQITE which produces a distribution of solution bitstrings to the problem. The bitstrings are the edge separators required to the partition the coarse graph into two subgraphs. If there is more than one level of nested dissection, the same procedure is repeated on the two subgraphs produced after the first level of nested dissection. The two subgraphs are once again coarsened and the graph partitioning is performed to produce more subgraphs. After the required number of levels of nested dissection are completed, the original matrix is reordered using the set of smaller graphs that have been produced by the partitioning. This reordered matrix has, hopefully, a better ``fill'' thereby reducing the number of non-zeros (memory requirement to store the matrix) and the number of computations required to perform the linear system solve for the FEA simulation. 

We estimate the resources required to solve the linear system by computing the number of non-zero fills arising from the \(LDL^T\) decomposition of the reordered matrix and the number of linear algebra computations required using symbolic factorization. We use these as the merit factors to compare the quality of the separators produced by the VarQITE optimization when compared with the internal classical heuristic LS-Gpart provided within LS-DYNA. The execution of the LS-DYNA tool is terminated after this computation. We use the LS-GPart heuristic as our classical benchmark in this work as it has demonstrated leading performance over a variety of problem geometries when compared to standard alternatives like METIS and Scotch~\cite{pellegrini1996scotch}.
We also compare the total wall clock time for the linear system solve using the VarQITE separators and the LS-GPart separators. In this case, we analyze the RoofCrush problem, the BloodPump problem, and the VibrationAnalysis problem. For each, we compute the total time for symbolic factorization, the actual factorization of the matrix and the time to solve the linear system. 

\subsection{Optimizing the VarQITE ansatz}\label{subsec:ansatz}

As described in Section~\ref{varQITE}, the VarQITE algorithm uses a parametrized quantum wavefunction to approximate the true wavefunction. The parametrized wavefunction can be instantiated using a quantum circuit ansatz consisting of parametrized quantum gates (1-qubit and 2-qubit quantum gates whose rotation angles are the parameters of the wavefunction). Exactly which gates are chosen and how they are ordered in the ansatz has a great impact on the convergence behavior of the VarQITE algorithm indicating that ansatz design is critical.

\begin{figure}[hbt]
 \centering   
    \includegraphics[width=\linewidth]{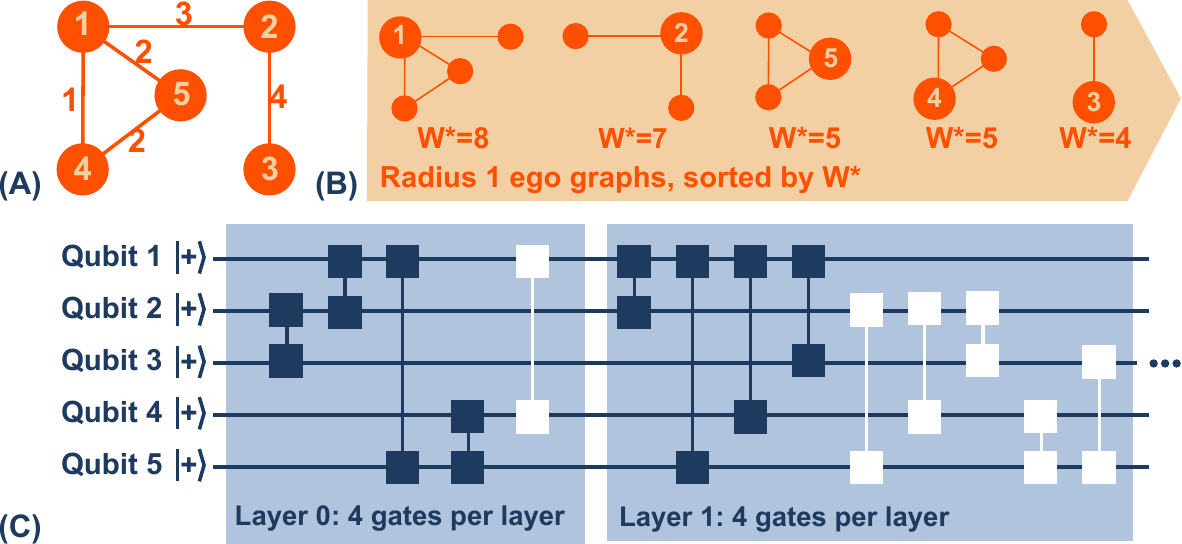}
    \caption{VarQITE ansatz: (A) An example of a $5$ node graph with the nodes and edge weights labeled, (B) The nodes sorted according to $W^*$, the total edge weight of their induced ego graph of radius $1$. (C) The VarQITE ansatz generated using the ego graphs. In this illustration, the ansatz is constructed with $4$ gates per layer (this number can be adjusted as required). The included gates are shown in dark blue while the excluded gates are shown in white (since there are only $4$ gates per layer, some of the gates will need to be excluded). The first layer (Layer 0) has entangling gates between qubits connected by the graph edges in the descending order of edge weights. The second layer (Layer 1) has entangling gates between qubits sorted by their induced ego graph of radius $1$. More layers may be added, generated by ego graphs of radii equal to $2, 3, 4$, etc. }
 \label{fig:heavyneighboransatz}
\end{figure}

We introduce the \texttt{HeavyNeighborsAnsatz}, a configurable layered ansatz that entangles certain qubits by exploiting graph structure. In this ansatz, every vertex of the graph is represented by a qubit and every entangling 2-qubit gate is represented by a $R_{ZY}(\theta)$ gate with a parametrized angle $\theta$. Given a graph $G$, a number of layers $\ell$, and an $\ell$-vector $\mathbf{g}$ determining the desired number of gates per layer, we construct the corresponding \texttt{HeavyNeighborsAnsatz} one layer at a time, as follows. At layer $0$, we sort the edges in the graph and append parametrized two-qubit gates on the pairs of qubits corresponding to the $\mathbf{g}_\ell$ heaviest edges. At layer $k > 0$, we sort the nodes in $G$ according to the total edge weight of their corresponding radius $k$ ego graph. Recall that the \textit{radius $r$ ego graph} of a node $v \in G$ is the induced subgraph centered at $v$ and including all nodes within a distance $r$ of $v$. Then we place parametrized two-qubit gates on the qubits corresponding to the nodes with the heaviest neighborhoods; in particular, if $\mathbf{s}^{(k)}$ denotes the indices of the nodes in $G$ sorted by decreasing radius $k$ ego graph weight, we place entangling gates on qubits $(\mathbf{s}^{(k)}_1, \mathbf{s}^{(k)}_2), (\mathbf{s}^{(k)}_1, \mathbf{s}^{(k)}_3), \ldots, (\mathbf{s}^{(k)}_1, \mathbf{s}^{(k)}_{\mathbf{g}_{\ell -k }})$.

Figure~\ref{fig:heavyneighboransatz} shows an example of a 5 qubit parametrized ansatz. All variational parameters in the ansatz are initialized to zero at the start of the VarQITE optimization, while the initial state $\ket{+}^{\otimes}$ is obtained from Hadamard gates applied to $\ket{0}^\otimes$ state. The VarQITE algorithm variationally optimizes the parameters of this ansatz to obtain the solutions to the graph partitioning problem. From a network flow perspective, the \texttt{HeavyNeighborsAnsatz} entangles qubits that correspond to nodes serving as gateways to regions where the accessible flow is large. These nodes have the highest impact on the cut size of a partition because they span the heaviest edges in the graph. The parametrized entangling gate across each such pair allows the ansatz to decide whether to increase the probability of observing one of the inputs in the opposite side of the partition.

In addition, the layered construction of the ansatz allows each candidate partition, encoded as a computational basis state in the superposition supported by the parametrized trial state, to be iteratively refined through a sequence of entangling gates as it goes through the layers. This is akin to the leading local classical heuristics like the Kernighan-Lin (KL) algorithm \cite{kernighan1970efficient} and the Fiduccia-Mattheyses (FM) algorithm \cite{fiduccia1988linear}. The refinement process is evident in the \texttt{HeavyNeighborsAnsatz}, as each layer incorporates increasingly global information. The first layer features the most ``local'' connectivity information based on direct connections between nodes, i.e. the edges in the graph. In contrast, the final layer places gates using ego graphs with the largest radius.

For the noiseless simulations described in Section~\ref{sec:noiseless_sims}, we used a 2-layer \texttt{HeavyNeighborsAnsatz}, with layer 1 consisting of all gates generated from the induced ego graph of radius $1$. Only gates that were already included in layer 0 were excluded in layer 1. This approach was chosen to enhance the expressibility of the ansatz and establish the baseline performance of the VarQITE algorithm. However, this technique requires a substantial number of two-qubit gates that makes it impractical for noisy NISQ-era QPU. Specifically, for a graph with $n$ nodes, the full ansatz would use $n (n{-}1)/2$ two-qubit gates, requiring the execution of $n^2{-}n{+} 1$ circuits on $n$ qubits, each with $n (n{-}1)/2$ two-qubit gates, at each VarQITE iteration to compute the expectation value of the hamiltonian and its gradients. Consequently, for the hardware runs described in Section~\ref{sec:hardware_runs}, we used an optimized ansatz in which we truncated the  \texttt{HeavyNeighborsAnsatz}, leading to a variant with a restricted number of gates per layer to reduce the total number of gates in the ansatz.

 \subsection{Evaluation of the merit factors}
 \label{subsec:merit-factor-eval}
 
\begin{figure}
    \centering
    \includegraphics[width=1.0\linewidth]{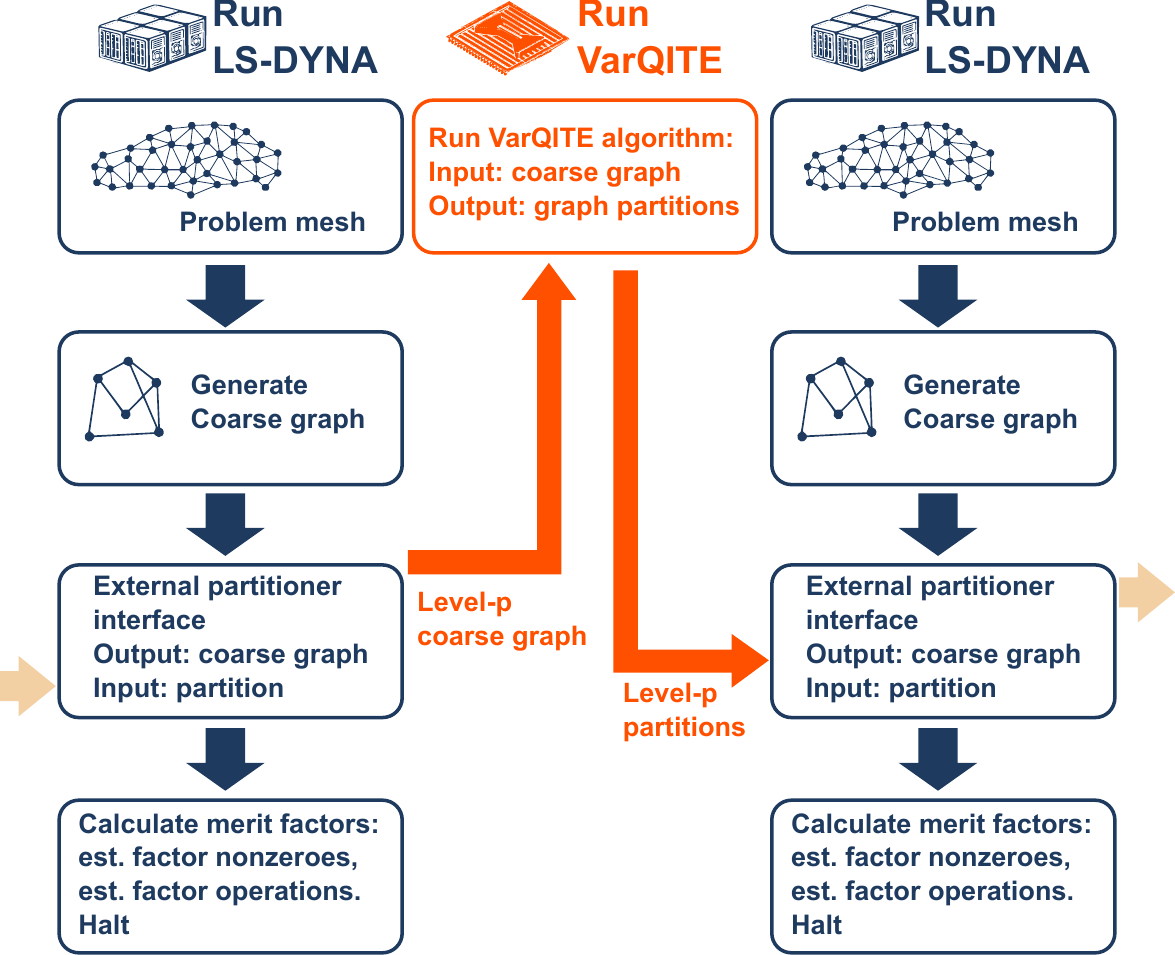}
    \caption{Integration of VarQITE into LS-DYNA for the evaluation of the factor of merit in a level-1 nested dissection process. LS-DYNA (orange arrows and boxes) executes two times in this integration framework. The VarQITE algorithm (blue arrows and boxes), which computes optimal partitions, operates externally between the two executions of LS-DYNA.}
    \label{fig:LS-DYNA_manual_pipeline}
\end{figure}

To assess the value of the VarQITE algorithm's solution, it is essential to incorporate it into the LS-DYNA workflow.
We developed a modified LS-DYNA version that enables an external subroutine call for graph partitioning, as shown in the red blocks of Figure~\ref{fig:LS-DYNA_manual_pipeline}. Through three separate runs, LS-DYNA first extracts the coarsened graph, then the VarQITE algorithm partitions it, and finally, LS-DYNA calculates the merit factors (through symbolic factorization) of the given partition which is the number of extra nonzeros created by ``fill-in'' and the number of operations required to the solve the linear system. This process allows for a direct performance comparison between LS-GPart and the VarQITE algorithm.

As shown in Figure~$\ref{fig:LS-DYNA_manual_pipeline}$, the procedure used to evaluate the partition for a single-level (level-1) nested dissection involves running LS-DYNA until the coarse graph is generated, at which point the program halts and the coarse graph is extracted. The VarQITE algorithm then solves the GPP on the extracted coarse graph. When the VarQITE algorithm converges, the quantum circuit is measured many times (shots) to obtain a sampled distribution of solutions from the underlying probability distribution of the quantum wavefunction. Notably, the VarQITE algorithm's strength lies in its ability to generate this sampled distribution which is focused on a
collection of low objective function values which may include the optimal solution. This allows for the exploration of merit factors from multiple near-optimal solutions, rather than being limited to a single value. 

This integrated process extends to multiple levels of nested dissection. To evaluate level-$L$ nested dissection, LS-DYNA is executed a total of $L+1$ times. On the $l$-th level of nested dissection, the VarQITE algorithm is executed to compute partitions for the $2^{l-1}$ coarse graphs extracted from the $(l)$-th level of nested dissection. These computed partitions are then returned to LS-DYNA by the external partitioner to compute the $(l+1)$-th level coarse graphs and so on till all $L$ levels of nested dissection are completed and the final merit factors are obtained.

\subsection{Enhancing VarQITE algorithm with classical heuristic refinement}\label{subsec:post-process}
The graph partitioning problem has been extensively studied, with numerous classical heuristics developed over the years, including the Kernighan-Lin (KL) algorithm \cite{kernighan1970efficient}, the Fiduccia-Mattheyses (FM) algorithm \cite{fiduccia1988linear}, and reinforcement learning-based approaches \cite{gatti2022graph}. These methods find widespread applications in Finite Element Analysis, VLSI design, parallel computing, and other optimization tasks.

We propose enhancing the performance of the VarQITE algorithm executed on noisy quantum hardware by integrating it with a modified version of the FM algorithm. The original FM algorithm uses a local search strategy guided by a gain function to identify optimal partitions while maintaining a balanced node count between partitions. However, FM and its variants often struggle under stringent constraints like nodal-weight balance \cite{kucar2004hypergraph}.

To address this issue, we introduce a key modification to the FM algorithm that prioritizes removing nodes from the heavier partition, simultaneously improving edge cuts and nodal-weight balance. We leverage the solution distribution produced by VarQITE as an initial configuration for the modified FM algorithm, significantly improving convergence and solution quality. This hybrid quantum-classical approach provides efficient local refinement of VarQITE-generated partitions and acts as an effective error mitigation strategy, enhancing reliability in the NISQ era.

A detailed description, including the modified FM algorithm, definitions, and mathematical formulations, can be found in Supplementary Information Section~\ref{SI:FM+VQITE}.

\section{Results}\label{sec:results}

In this section, we first present the results from running VarQITE on a noiseless statevector simulator using the ansatz described in Section~\ref{subsec:ansatz}. This is followed by a comparison of the merit factors obtained from the GPP solutions produced by VarQITE and the classical heuristic in LS-DYNA. We then present the results from executing VarQITE on IonQ quantum hardware and the refinement of the GPP solutions using the modified FM algorithm. 

\subsection{Results of noiseless VarQITE simulations}\label{sec:noiseless_sims}

The two GPP problems of interest - RoofCrush and BloodPump were executed using VarQITE on a noiseless quantum simulator. The VarQITE optimization requires $2n$ gradient calculations of the quantum ansatz corresponding to n parameters at each iteration. For each circuit execution, $2,000$ shots were used to sample the quantum state. We explored graph coarsening of the two problems from 10 to 32 vertices, with each vertex mapped on to one qubit of the quantum circuit. We studied the performance of the VarQITE algorithm with one and two levels of nested dissection, denoted respectively Level-1 or L1, and Level-2 or L2. All simulations were run using the IonQ statevector simulator. Simulations for graphs with 24 or fewer nodes were executed in parallel on AMD EPYC 7763 compute nodes, while simulations for graphs with more than 24 nodes were executed on Nvidia A100 GPUs. 

\begin{figure}[hbt]
    \centering
    \includegraphics[width=1.0\linewidth]{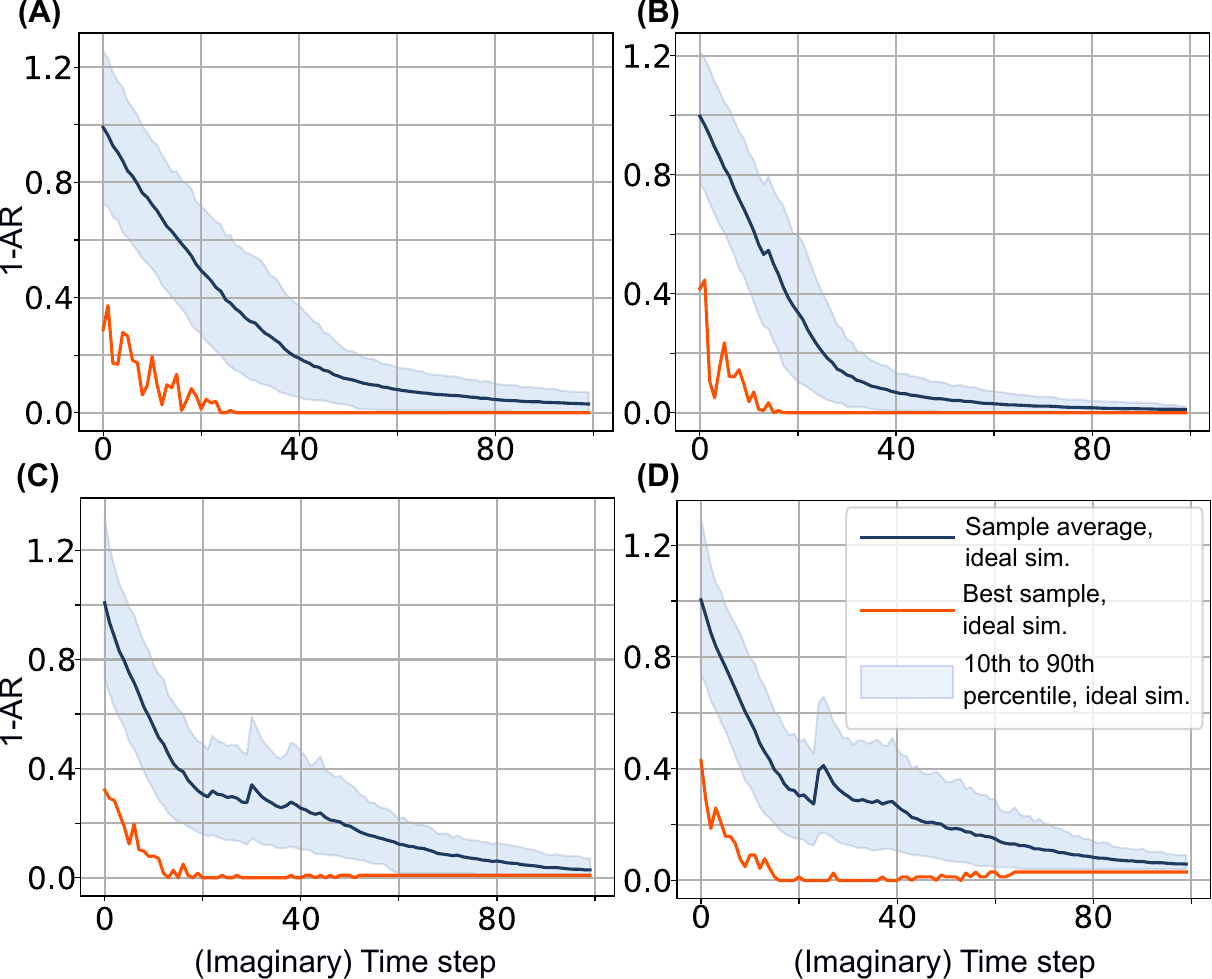}
    \caption{Convergence of VarQITE on a Noiseless Simulator: Plots for (A) RoofCrush 30 nodes, (B) RoofCrush 32 nodes, (C) BloodPump 30 nodes, (D) BloodPump 32 nodes. The y-axis shows the value of 1-AR where AR is the Approximation Ratio of the solution. The solid blue and purple lines correspond to the average and best value from the sampled distribution, respectively. The shaded blue region corresponds to the values between the 10th and 90th percentile of the sampled distribution. All results were obtained using 2,000 shots to sample the quantum circuit.}
    \label{fig:baseline_convergence}
\end{figure}

We define the Approximation Ratio as the energy of the solution divided by the energy of the optimal solution, where the energy of a solution is obtained by evaluating the QUBO hamiltonian defined in Section~\ref{sec:HQUBO} on the solution. The value of 1-AR should decrease as the VarQITE algorithm evolves, because the AR of the solutions found would improve towards to optimal solution. A value of 0.0 on the y-axis is reached when the optimal solution is sampled.

Figure~\ref{fig:baseline_convergence} shows the convergence behavior of the VarQITE algorithm for the RoofCrush and the BloodPump GPP problems. The y-axis represents the value of 1 minus the Approximation Ratio (AR) of the solutions found at a given iteration of VarQITE. At each iteration the measurement of the quantum circuit produces 2,000 binary bitstrings, which represent potential partitions. The best solution sampled at any given iteration is shown in orange. The average value over the entire sampled distribution is depicted in dark blue, while the lighter blue band indicates the range of values between the 10th and 90th percentiles of sampled distribution.
As the VarQITE algorithm iteratively amplifies the probability of states with lower energy, the initial distribution of solutions (which is broad and uniform) rapidly narrows, as shown by the gap between the 10th and 90th percentiles becoming increasingly smaller over successive iterations. Ultimately, the distribution peaks over the optimal solution and other near-optimal solutions. This convergence behavior is observed consistently in the plots presented and the optimal solution is found in each case. Figure~\ref{fig:RC_histogram}(A) in the the Supplementary Information Section~\ref{SI:FM+VQITE} also shows the final sampled probability distribution for the RoofCrush problem with 32 vertices.

\subsection{Comparing merit factors between VarQITE and LS-GPart}
\label{sec:merit-figures}

The estimated number of non-zeroes in the triangular factor and the estimated number of operations required to factor the matrix during the symbolic factorization step were the figures of merit used in this analysis. These metrics, which are influenced by the provided partition, are crucial parameters affecting the total time to solution of FEA problem. In this section, we compare the merit factors produced by using the VarQITE partitioner against LS-DYNA internal heuristic (LS-GPart). The VarQITE algorithm's ultimately goal is to reduce the overall time to solve each linear system. 

We limited the comparison to coarse graphs ranging from 10 to 32 nodes to match the capabilities of our quantum simulator. We also used LS-GPart to compute the merit factors for graphs coarsened up to 50,000 nodes to get a sense of the performance of the system in the regime typically used for standard LS-DYNA production runs. When the VarQITE algorithm converges, it produces a peaked distribution over the optimal/near-optimal GPP solutions. We investigated the merit factors for several of these near-optimal solutions in addition to the optimal solution. 

The solutions found by VarQITE aim to minimize the total objective function, which is a combination of edge cut cost and nodal weight balance. We limit the evaluation of the merit factor to solutions that preserve the nodal weight balance within a 5\% tolerance in order to maintain nodal weight balance as much as possible. The primary goal of this restriction is to allow for appropriate load balancing when LS-DYNA solves the actual linear system in parallel on multiple compute nodes. The weight balance is a heuristic hyperparameter, and more research is necessary to understand how it affects the total time to solution of the linear system.

\begin{figure}[h!]
    \centering
    \includegraphics[width=1.0\linewidth]{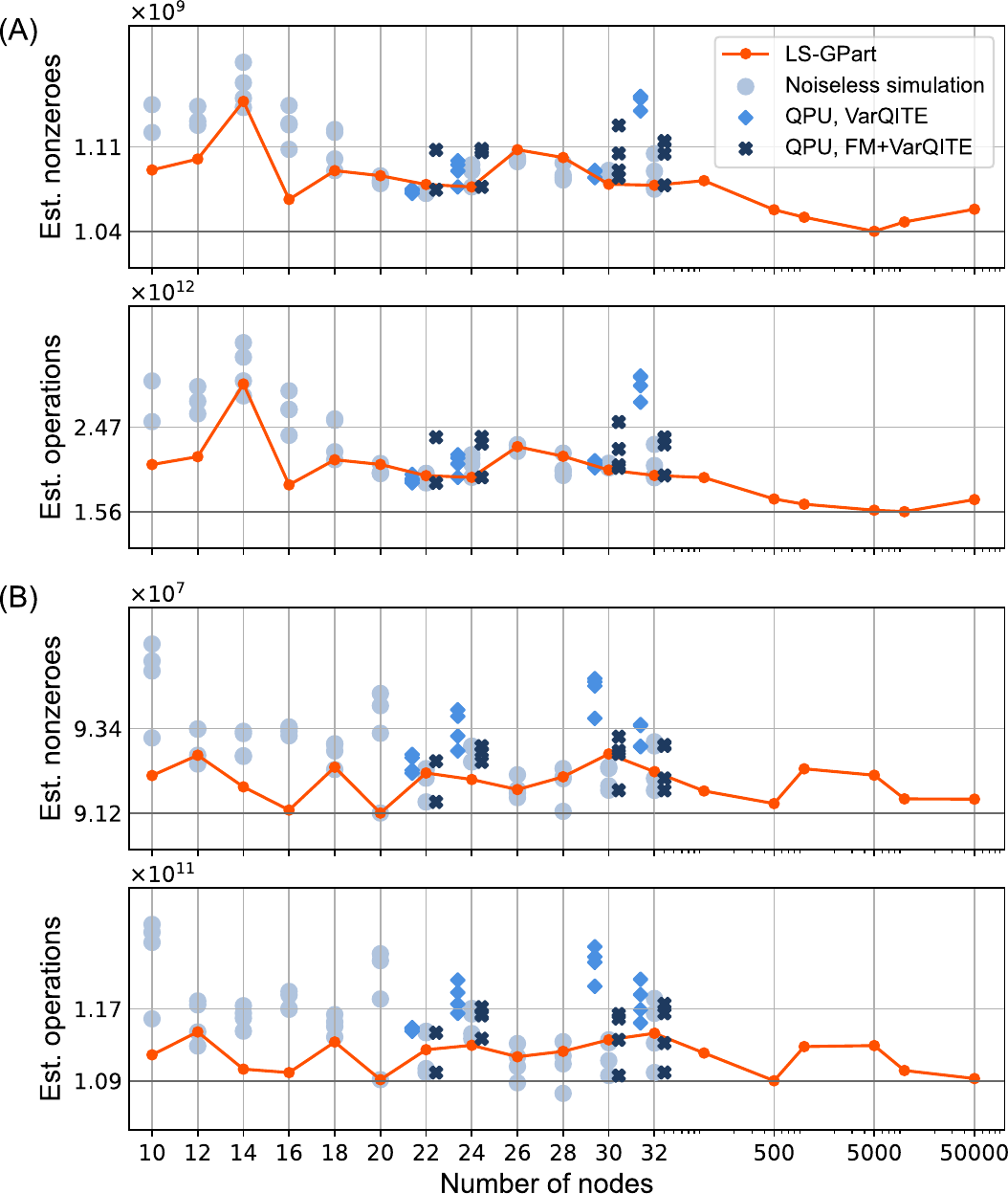}
    \caption{Comparison of merit factors resulting from the VarQITE algorithm and LS-GPart. The two panels shown are (A) RoofCrush and (B) BloodPump, and each problem uses a Level 1 nested dissection. Each panel includes (top) the number of non-zeros (``fill-in''), and (bottom) the number of operations estimated from symbolic factorization by LS-DYNA. The VarQITE data is restricted to graph partitions that maintain a nodal weight balance within a 5\% tolerance. The QPU data is from graphs with the same number of nodes as the corresponding simulation data, the plotting is offset for clarity. The plot also shows results from applying the modified FM algorithm to the QPU data. Shown on the horizontal axis are different numbers of nodes for the coarsened graph, corresponding to the number of qubits. Note that the largest VarQITE experiments reported on have 32 nodes/qubits, however, the horizontal scale is extended logarithmically to much larger sizes of coarsening in order to capture the regime in which LS-GPart operates. The 4 best solutions are plotted for each number of nodes/qubits in the coarsening, shown as the grey dots in the charts. As can be seen in the lower panel, for the BloodPump problem, solutions obtained by coarsening to 28 nodes lead to better merit factors when compared to LS-GPart (orange dots and lines) in terms of non-zeroes and in terms of factorization operations, even when compared to numbers of nodes that are as high as 50,000.}
    \label{fig:merit_factors-L1}
\end{figure}

Figure~\ref{fig:merit_factors-L1} shows the merit factors obtained from a Level-1 nested dissection compared against LS-GPart. The VarQITE algorithm provides solutions with merit factors that are either equal or lower than LS-GPart solutions for graphs with 18 or more nodes. This is the case for both problem instances. Note that for graphs with a small number of nodes ($\leq$ 16), LS-GPart produces solutions exceeding the 5\% nodal weight balance threshold; this is likely due to the limited options available. Because VarQITE solutions are restricted to those within the nodal weight tolerance, the merit factors from VarQITE are higher than LS-GPart in these cases. The decreasing trend in average merit factors as the number of nodes in the graph increases suggests that the VarQITE algorithm can generate solutions with even lower merit factors when applied to significantly larger graphs. Notably, for the BloodPump problem (Figure~\ref{fig:merit_factors-L1}(B)) the VarQITE merit factors for the 26 and 28 nodes graph instances surpasses those from LS-GPart, even for graphs with 50,000 nodes. This suggests that certain problem types with specific geometries may exist for which the improvement in merit factors plateaus with increasing number of nodes in the coarsened graph. This demonstrates an intriguing potential for the VarQITE algorithm to provide a benefit for such specific problem types, even in the near term. Investigation of the existence and structure/properties of such problems will require further research.

It is important to note that the QUBO objective function value of the solutions are not perfectly correlated with the merit factor. This is due to several factors, including the coarsening of the graph to a small number of nodes and the use of heuristics in the partitioning and matrix reordering processes. This holds true even when the graph is coarsened to the standard production setting of 10,000--50,000 nodes. In order to obtain the best solution to the overall matrix reordering problem, the merit factor of a set of near-optimal solutions may be computed via symbolic factorization (which is efficiently performed classically) so that the solution with the best merit factor may be selected as the graph partitioner. The potential advantage of using VarQITE in this context lies in its ability to produce a peaked distribution over multiple near-optimal solutions to the GPP, since these solutions may achieve lower merit factors. As the quantum hardware advances and the qubit counts increase, the benefit of using VarQITE for graph partitioning in terms of reducing overall FEA solution time becomes an intriguing possibility. 

We also extended the LS-DYNA nested dissection workflow to level 2 as explained in Figure~\ref{fig:LS-DYNA_manual_pipeline}. The results of these simulations are available in the Supplementary Information (SI) \ref{SI:FM+VQITE}.

\subsection{Measuring total wall clock time for the linear solver}\label{sec:wct}

In addition to the merit figure computations obtained in Section~\ref{sec:merit-figures}, we conducted  experiments to investigate how the partitions produced by our VarQITE-powered GPP solver impact the wall clock time (WCT) required by LS-DYNA to execute a linear system solve. For these simulations we used the RoofCrush and BloodPump problems from the previous section, as well as the VibrationAnalysis problem from Table~\ref{tab:problem_instances}. Every experiment was conducted on the same machine, built with $4$ Intel Xeon Gold 6242 CPUs, 16 cores each, and 1.5TB of memory to ensure consistency across experiments. We vary the number of MPI ranks (processes which collaborate to run the simulation) from 8 to 24.
For the BloodPump problem, the total WCT measured was too small to observe a consistent trend when run on a larger number of MPI ranks. We therefore, refined the previously used mesh to increase the number of vertices to 2.6 million and the number of edges to 40 million to ensure that the measured WCT was large enough to discern a consistent trend.

\begin{figure*}[hbt]
    \centering
    \includegraphics[width=\linewidth]{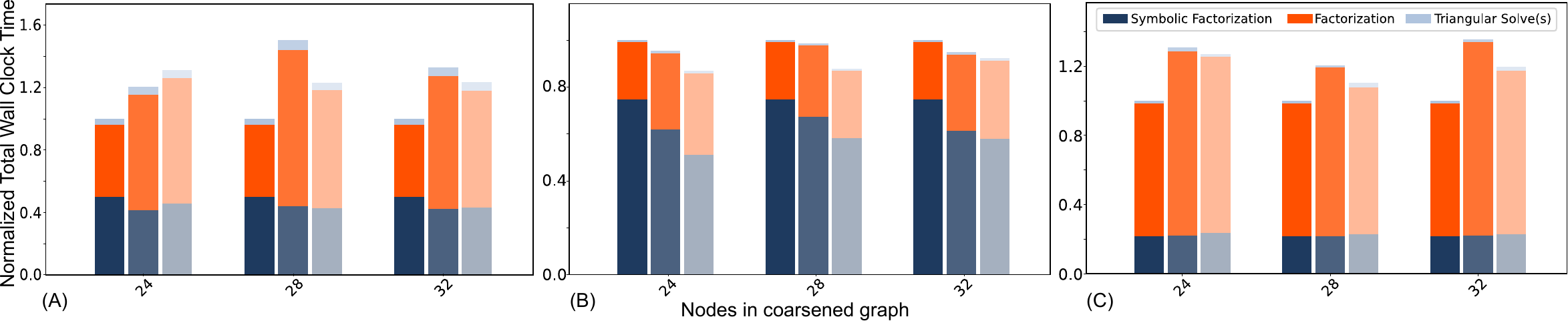}
    \caption{Total wall clock time (WCT) comparison for linear system solve of the (A) RoofCrush problem on 8 MPI ranks, (B) BloodPump problem on 24 MPI ranks and (C) VibrationAnalysis problem on 16 MPI ranks. All simulations were done on a 4 Intel Xeon Gold 6242 CPUs, 64 cores in total; and 1.5TB memory. In each group, the leftmost bar represents total WCT when coarsening the graph to 10,000 nodes (production setting in LS-DYNA/LS-GPart). The middle and right bars compare WCT when coarsening the graph to different numbers of nodes shown on the horiontal axis, and using the internal (LS-GPart) vs. the external (varQITE-based) partitioner, i.e., the bars on the right show the quantum-enhanced performance. Note that the external partitioner leads to shorter WCT for all experiments shown in this figure. Further, there is a trend in case of RoofCrush in that the wall clock times seem to decrease with increasing number of qubits. Finally, in case of BloodPump, the WCT of the solver resulting from the varQITE partitioner is lower than the WCT of the production solver. The differences of the left to the right bar in panel (B) are, reading from left to right, $-11.93\%$, $-11.51\%$, and $-9.80\%$. This means that the quantum solution improved over the industry-grade heuristic by close to $12\%$.}
    \label{fig:wct_expts}
\end{figure*}

Figure~\ref{fig:wct_expts} shows the comparison of the total WCT required for a linear system solve when using the optimal reordering computed using LS-GPart and the VarQITE-powered GPP solver. For the WCT using the VarQITE solver, the partition which yielded the lowest WCT amongst the 10 lowest energy partitions from the solver's optimized distribution is plotted. The figures explore how the total WCT scales as a function of the number of nodes in the coarsened graph. The plot compares the resulting WCTs normalized with respect to the WCT required by the reference code, which is LS-DYNA configured with its production settings (graphs coarsened to 10,000 nodes). In order to obtain realistic results, we distributed these computations across 8, 16, or 24 MPI ranks, much like a typical LS-DYNA user. For these experiments, we refined the mesh of the BloodPump problem by making it 100x more dense so the factorization time was meaningful enough to measure improvements.
 
The figures provide evidence that our VarQITE-powered partitioner improves upon LS-GPart in most cases when used on the same coarsened graph. This shows that the VarQITE algorithm is able to provide partitions that improve the WCT across different FEA models displaying the applicability of the algorithm to a spectrum of different problem types. Figure~\ref{fig:wct_expts}(A) illustrates a decreasing trend in WCT for the RoofCrush problem as we increase the number of nodes in the coarsened graph. The same trend can be noticed in Figure~\ref{fig:wct_expts}(C) for the VibrationAnalysis problem (although there is a little more noise in this case). The decreasing trend provides confidence that the VarQITE algorithm could be scaled to much larger graph sizes and may continue to improve the WCT, matching or even improving upon the production settings (graphs coarsened to 10,000 nodes) in the near future. In addition, Figure~\ref{fig:wct_expts}(B) suggests there are some problem geometries where we may find comparable WCT when coarsening to a few dozen nodes as we do when coarsening to the production-level of 10,000 nodes. This result is quite encouraging as a stepping-stone towards quantum utility, as it suggests there could be certain problem types where we may find early advantage when using a VarQITE-powered partitioner over leading existing classical methods. Much more work is required in this area to identify these problem types and develop an understanding of why they may be amenable to such advantages.

\subsection{Hardware experiments}\label{sec:hardware_runs}

\begin{figure}[hbt]
\includegraphics[width=1.0\linewidth]{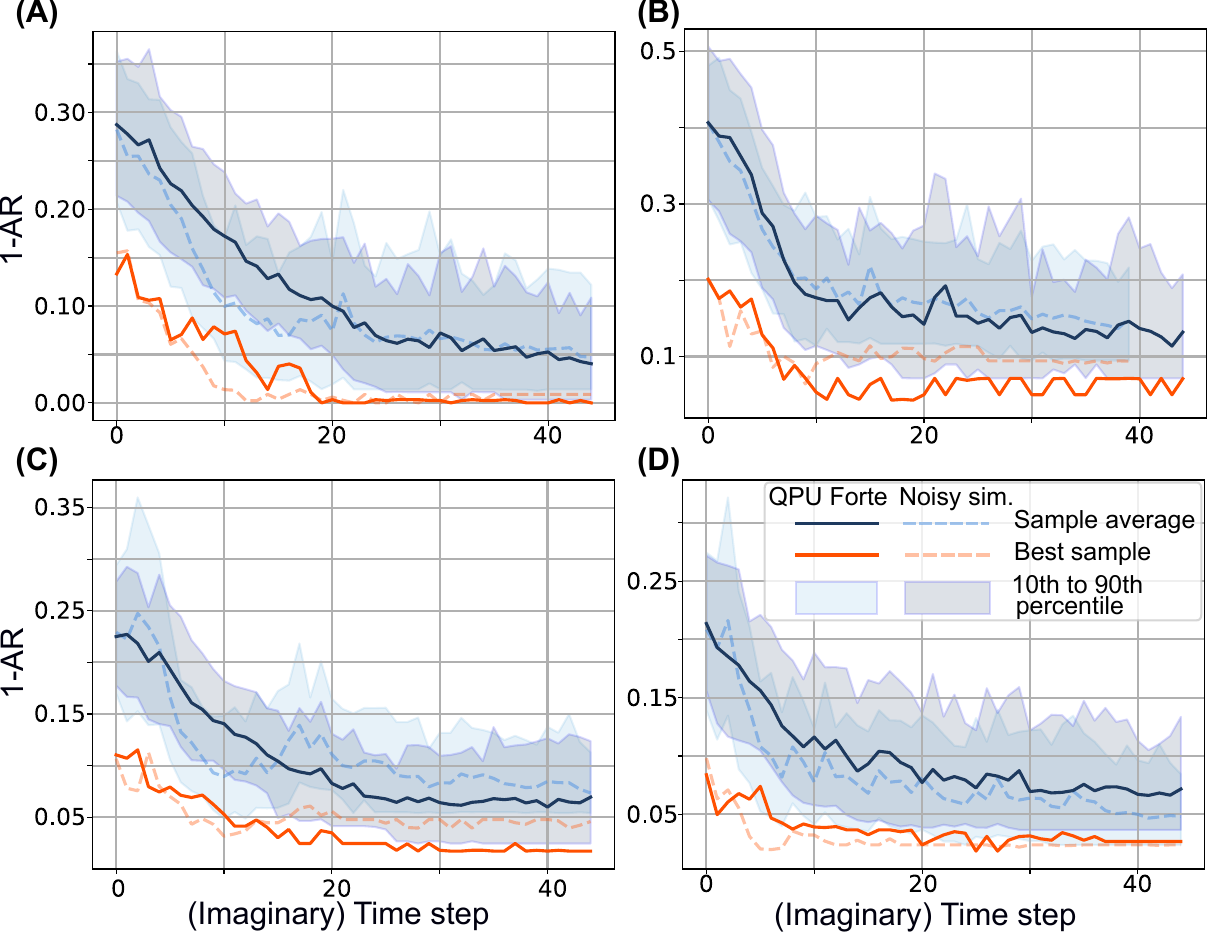}
\caption{Plots of convergence of VarQITE on IonQ QPU Forte, (A) RoofCrush 30 nodes, (B) RoofCrush 32 nodes, (C) BloodPump 30 nodes, (D) BloodPump 32 nodes. The y axis shows the value of 1-AR where AR is the approximation ratio of the solution. The solid blue line and solid purple line correspond to the average and best value sampled on IonQ Forte. The dashed green line and dashed purple line correspond to the average and best value sampled using the IonQ noisy simulator. The shaded blue and green regions correspond to the values of solutions between the 10th and 90th percentile in the sampled distribution.}
\label{fig:forte_convergence}
\end{figure}

A few select graph instances from the RoofCrush and BloodPump problems were executed on IonQ quantum processing units (QPUs) Aria and Forte. The IonQ Aria QPU uses up to 25 addressable Ytterbium (Yb) ions linearly arranged in an ion trap, while the IonQ Forte QPU has 36 addressable Ytterbium (Yb) ions. Qubit states are implemented by utilizing two states in the ground hyperfine manifold of the Yb ions. Manipulating the qubits in the Aria and Forte QPUs is done using mode-locked 355nm laser pulses, which drive Raman transitions between the qubit states. By configuring these pulses, arbitrary single-qubit gates and Mølmer-Sørenson type two-qubit gates \cite{sorensen1999quantum} can both be realized. As of this publication, the Aria QPU has demonstrated performance at the level of 25 algorithmic qubits and the Forte QPU has demonstrated performance at the level of 36 algorithmic qubits \cite{IonQ_blog, IonQ_Aria, IonQ_Forte, lubinski2023application}. In order to mitigate the effect of systematic errors on the Aria/Forte QPU, error mitigation via symmetrization was used \cite{maksymov2023enhancing}. After executing multiple circuit variants with distinct qubit to ion mappings and twirling, the measurement statistics was aggregated using component-wise averaging.

Graph partitioning problems from both the RoofCrush and BloodPump problem instances with 22 and 24 nodes were executed on IonQ Aria and graph partitioning problems with 30 and 32 nodes were executed on IonQ Forte. The graph problems were chosen to demonstrate the performance of the VarQITE algorithm and the IonQ QPUs as a function of problem size as well as the capability to handle different types of graphs for a given problem size. To execute VarQITE algorithm on the QPU, the \texttt{HeavyNeighborsAnsatz} described in \ref{subsec:ansatz} was used with  60 parametrized $R_{ZY}$ gates for the RoofCrush problem instance and 82 parametrized $R_{ZY}$ gates for the BloodPump problem instance. The number of gates in the ansatz was chosen taking into account the total QPU execution time for the algorithm and the fact that the ansatz does not necessarily require gates across all pairs of qubits to solve these graph problems. The number of two-qubit gates selected in the truncated ansatz balances the ansatz expressibility, the deleterious effects of QPU noise as more gates are added, and the total execution time on the QPU.

 All measurements of the quantum circuits were performed in the computational basis with 128 shots per circuit evaluation and measurement. The hybrid quantum optimization algorithm was executed entirely on the IonQ QPU hardware from initialization until convergence. The evolution of the average error (defined as 1-AR) as a function of the iterations of the VarQITE algorithm is shown in Figure~\ref{fig:forte_convergence} for graphs of sizes 30 and 32 from both the RoofCrush and BloodPump problem instances executed on IonQ Forte. The average error decreases steadily as the higher energy eigenstates of the hamiltonian are projected out and the final distribution peaks on/near the optimal solution. This demonstrates the robustness of the VarQITE algorithm to both shot noise (128 shots per circuit measurement) and QPU noise along with its excellent performance on different types of graphs with differently weighted edges and connectivity. In all cases, the algorithm is able to attain an average error rate of < 20\% and a best sample error rate of < 5\%. Figure~\ref{fig:RC_histogram}(B) in the Supplementary Information Section~\ref{SI:FM+VQITE}  shows the sampled distribution from the last iteration of the VarQITE algorithm for the RoofCrush GPP problem with 30 nodes executed on IonQ Forte. 

\begin{table*}[hbt]
\centering
\resizebox{\linewidth}{!}{%
    \setlength{\tabcolsep}{3pt} 
    \begin{tabular}
    {cc@{\hskip 24pt}ccc@{\hskip 24pt}ccc@{\hskip 24pt}ccc}
    \toprule
     & & \multicolumn{3}{@{}c}{RoofCrush\phantom{xxxxxx}} & \multicolumn{3}{@{}c}{BloodPump\phantom{xxxxxx}} & \multicolumn{3}{@{}c}{VibrationalAnalysis\phantom{x}} \\ [1ex]
    MPI ranks & Nodes & WCT (s) & $\Delta$ (\%) & $\Delta_{\text{prod}}$ (\%) & WCT (s) & $\Delta$ (\%) & $\Delta_{\text{prod}}$ (\%) & WCT (s) & $\Delta$ (\%) & $\Delta_{\text{prod}}$ (\%) \\
    \midrule
    8 & 24 & 27.51 & -10.30 & 52.06 & 76.80 & -4.07 & -2.73 & 857.83 & -41.35 & 100.94 \\
    8 & 28 & 24.87 & -13.83 & 37.50 & 72.73 & 2.31 & -7.88 & 896.25 & -44.11 & 109.94 \\
    8 & 32 & 24.23 & -14.18 & 33.92 & 68.15 & 4.54 & -13.69 & 886.20 & -20.74 & 107.58 \\
    16 & 24 & 14.99 & 8.36 & 18.59 & 59.38 & -17.46 & 6.69 & 323.65 & -2.94 & 30.73 \\
    16 & 28 & 18.20 & -11.74 & 43.95 & 51.16 & -3.74 & -8.07 & 298.61 & -8.58 & 20.61 \\
    16 & 32 & 15.09 & 7.54 & 19.38 & 48.12 & 2.66 & -13.55 & 335.79 & -11.79 & 35.63 \\
    24 & 24 & 11.63 & 8.60 & 20.71 & 42.32 & -8.86 & -4.65 & 232.78 & -6.49 & 40.43 \\
    24 & 28 & 14.49 & -18.27 & 50.38 & 43.80 & -11.17 & -1.32 & 208.32 & -1.77 & 25.67 \\
    24 & 32 & 12.79 & -7.01 & 32.76 & 42.13 & -2.69 & -5.09 & 185.01 & 31.85 & 11.61 \\
    \bottomrule
    \end{tabular}
 }
\caption{Total WCT for RoofCrush, BloodPump and VibrationalAnalysis simulations. This table details the results illustrated in Figure \ref{fig:wct_expts}; here, the WCT column reports the total wall clock time spent solving the associated linear system using the graph partition computed by our VarQITE algorithm, and the $\Delta$ column compares that WCT against the time needed to solve the same linear system when coarsening to the same degree, and the $\Delta_\mathrm{prod}$ column compares it to the total time required to solve the same linear system using LS-DYNA's production settings. We observe improvements over the production settings only in the BloodPump CFD calculation.}
\end{table*}

The merit factors computed from the solutions obtained from quantum hardware are shown in Figure~\ref{fig:merit_factors-L1}. The hardware merit factors are close to the simulation merit factors showing the high quality of solutions found on IonQ hardware and the robustness of the VarQITE algorithm. We observe that for the largest problem sizes of 30 and 32 qubits, the hardware merit factors are higher than the simulation merit factors due to QPU noise and shot noise. With further improvements in the fidelity of quantum hardware, solutions with lower merit factors can be found. In order to improve the merit factors of the solutions, the modified FM algorithm was used to refine the solutions by taking the solutions obtained from the hardware runs as input. These are shown in Figure~\ref{fig:merit_factors-L1} as ``FM+VarQITE''. The merit factors from these solutions match the best simulation data and outperform the LS-DYNA classical heuristic for many problem instances, e.g., for the BloodPump problem. The modified FM algorithm can thus be used as a means of error mitigation on noisy QPU results to recover high quality solutions with low merit factors. This demonstrates the high performance and robustness of the combined ``VarQITE+modified FM'' approach making the use of near term quantum hardware to deliver tangible enhancements to certain large scale industrial problems a possibility in the near future.

\section{Conclusions and Outlook}\label{sec:future-outlook}

We integrated the VarQITE quantum algorithm into an industrial workflow for solving large scale problems in FEA. This shows that VarQITE may be used to solve graph partitioning problems arising in the solution of large sparse linear systems in various domains from mechanical engineering to computational fluid dynamics. We showed that VarQITE can find partitions that improve the WCT across different FEA models when compared to a classical heuristic on the same graphs. These results provide confidence that the VarQITE algorithm could be scaled to much larger graph sizes and would continue to improve the WCT, matching and even improving upon the production settings in the near future as quantum computers continue to scale.

We have also demonstrated full hybrid on-QPU optimization of the graph partitioning problem for graphs of varying types, connectivity and sizes on IonQ Aria and IonQ Forte QPUs. Our results show that the VarQITE algorithm is capable of finding good solutions to the GPP for graphs coarsened up to 32 nodes with convergence to high solution probability even on noisy quantum hardware indicating the robustness of the quantum algorithm against shot noise and QPU noise. The high quality of results (especially when combined with classical heuristic refinement) obtained for the different types of problems studied suggests the robustness of the algorithm in solving generic instances of the graph partitioning problem arising in industrially relevant large scale FEA problems. 

 Our results show that there are some problem geometries where we may find improvements to WCT when coarsening to a few dozen nodes as we do when coarsening to the industrial-level of 10,000 nodes. We find that WCT by up to 12\% for some problems such as a blood pump model and that for other models there seems to be a trend that shows improvements with larger number of qubits. 
 
 Future work will explore problems that may be amenable to quantum enhancements. An example is vibrational analysis, where the problem requires the determination of the lowest eigenmodes of the matrix. Eigensolvers such as shift-and-invert Lanczos~\cite{grimes1994shifted} involve a sequence of factorizations and backsubstitutions where only the values of the input matrix change, not the nonzero pattern; therefore, the reordering need only be done once at the beginning of the simulation. 
 
 A superior reordering of the matrix provided by a better solution from VarQITE can potentially have a much larger impact in the total WCT for such problems. The identification of such problems should become the focus of investigation of potential quantum commercial utility in the near term as quantum hardware continues to scale and improve. The low circuit depth of our ansatz, the minimal resource requirements of the VarQITE algorithm, superior empirical performance, and scalability of our approach make it a promising candidate for solving interesting industrial-scale problems in the future. 

Future work will also investigate further the classical heuristic inspired by Fiduccia-Mattheyses to improve the quality of VarQITE solutions obtained from hardware runs. 



\begin{thebibliography}{10}
\providecommand{\url}[1]{#1}
\csname url@samestyle\endcsname
\providecommand{\newblock}{\relax}
\providecommand{\bibinfo}[2]{#2}
\providecommand{\BIBentrySTDinterwordspacing}{\spaceskip=0pt\relax}
\providecommand{\BIBentryALTinterwordstretchfactor}{4}
\providecommand{\BIBentryALTinterwordspacing}{\spaceskip=\fontdimen2\font plus
\BIBentryALTinterwordstretchfactor\fontdimen3\font minus \fontdimen4\font\relax}
\providecommand{\BIBforeignlanguage}[2]{{%
\expandafter\ifx\csname l@#1\endcsname\relax
\typeout{** WARNING: IEEEtran.bst: No hyphenation pattern has been}%
\typeout{** loaded for the language `#1'. Using the pattern for}%
\typeout{** the default language instead.}%
\else
\language=\csname l@#1\endcsname
\fi
#2}}
\providecommand{\BIBdecl}{\relax}
\BIBdecl

\bibitem{yannakakis1981computing}
M.~Yannakakis, ``Computing the minimum fill-in is {NP}-complete,'' \emph{SIAM Journal on Algebraic Discrete Methods}, vol.~2, no.~1, pp. 77--79, 1981.

\bibitem{george1973nested}
A.~George, ``Nested dissection of a regular finite element mesh,'' \emph{SIAM journal on numerical analysis}, vol.~10, no.~2, pp. 345--363, 1973.

\bibitem{METIS}
``{METIS}: A software package for partitioning unstructured graphs, partitioning meshes, and computing fill-reducing orderings of sparse matrices,'' technical report, Department of Computer Science, University of Minnesota, 1998.

\bibitem{andreev2006gpp}
K.~Andreev and H.~Racke, ``Balanced graph partitioning,'' \emph{Theory of Computing Systems}, vol.~39, no.~6, p. 929–939, Oct 2006.

\bibitem{Morris2024-gj}
\BIBentryALTinterwordspacing
T.~D. Morris, A.~Kaushik, M.~Roetteler, and P.~C. Lotshaw, ``Performant near-term quantum combinatorial optimization,'' \emph{arXiv [quant-ph]}, 24~Apr. 2024. [Online]. Available: \url{http://arxiv.org/abs/2404.16135}
\BIBentrySTDinterwordspacing

\bibitem{Ansys_LSDYNA}
``Ansys {LS-DYNA},'' https://www.ansys.com/products/structures/ansys-ls-dyna.

\bibitem{Yuan2019-oj}
\BIBentryALTinterwordspacing
X.~Yuan, S.~Endo, Q.~Zhao, Y.~Li, and S.~C. Benjamin, ``Theory of variational quantum simulation,'' \emph{Quantum}, vol.~3, no. 191, p. 191, Oct. 2019. [Online]. Available: \url{https://quantum-journal.org/papers/q-2019-10-07-191/pdf/}
\BIBentrySTDinterwordspacing

\bibitem{McArdle2019-zp}
\BIBentryALTinterwordspacing
S.~McArdle, T.~Jones, S.~Endo, Y.~Li, S.~C. Benjamin, and X.~Yuan, ``Variational ansatz-based quantum simulation of imaginary time evolution,'' \emph{Npj Quantum Inf.}, vol.~5, no.~1, pp. 1--6, Sep. 2019. [Online]. Available: \url{https://www.nature.com/articles/s41534-019-0187-2}
\BIBentrySTDinterwordspacing

\bibitem{Mitarai2018-mu}
\BIBentryALTinterwordspacing
K.~Mitarai, M.~Negoro, M.~Kitagawa, and K.~Fujii, ``Quantum circuit learning,'' \emph{Phys. Rev. A}, vol.~98, no.~3, p. 032309, Sep. 2018. [Online]. Available: \url{http://link.aps.org/pdf/10.1103/PhysRevA.98.032309}
\BIBentrySTDinterwordspacing

\bibitem{Schuld2019-cq}
\BIBentryALTinterwordspacing
M.~Schuld, V.~Bergholm, C.~Gogolin, J.~Izaac, and N.~Killoran, ``Evaluating analytic gradients on quantum hardware,'' \emph{Phys. Rev. A}, vol.~99, no.~3, p. 032331, Mar. 2019. [Online]. Available: \url{http://link.aps.org/pdf/10.1103/PhysRevA.99.032331}
\BIBentrySTDinterwordspacing

\bibitem{del2020performance}
F.~Del Pin, C.-J.~Huang, I.~Çaldichoury, \& R.~Paz, ``On the performance and accuracy of PFEM-2 in the solution of biomedical benchmarks'', \emph{ Computational Particle Mechanics}, vol.~7, pp. 121--138 (2020).

\bibitem{marzougui2014development}
D.~Marzougui, D.~Brown, H.~K.~Park, C.-D.~Kan, \& K.~S.~Opiela, ``Development \& validation of a finite element model for a mid-sized passenger sedan'', \emph{13th International LS-DYNA Users Conference}, 2014.

\bibitem{reichert2016validation}
R.~Reichert, P.~Mohan, D.~Marzougui, C.-.D.~Kan, \& D.~Brown, ``Validation of a Toyota Camry Finite Element Model for Multiple Impact Configurations'', \emph{SAE Technical Paper 2016-01-1534}, 2016.

\bibitem{singh2016update}
H.~Singh, C.-D.~Kan, D.~Marzougui, \& R. M. Morgan, ``Update to Future Midsize Lightweight Vehicle Findings in Response to Manufacturer Review and IIHS Small-Overlap Testing'', DOT Technical Report DOT HS 812 237, 2016.

\bibitem{pellegrini1996scotch}
F.~Pellegrini \& J.~Roman, Scotch: ``A software package for static mapping by dual recursive bipartitioning of process and architecture graphs'', \emph{High-Performance Computing And Networking}, pp. 493--498 (1996).

\bibitem{kernighan1970efficient}
B.~W. Kernighan and S.~Lin, ``An efficient heuristic procedure for partitioning graphs,'' \emph{The Bell system technical journal}, vol.~49, no.~2, pp. 291--307, 1970.

\bibitem{fiduccia1988linear}
C.~M. Fiduccia and R.~M. Mattheyses, ``A linear-time heuristic for improving network partitions,'' in \emph{Papers on Twenty-five years of electronic design automation}, 1988, pp. 241--247.

\bibitem{gatti2022graph}
A.~Gatti, Z.~Hu, T.~Smidt, E.~G. Ng, and P.~Ghysels, ``Graph partitioning and sparse matrix ordering using reinforcement learning and graph neural networks,'' \emph{Journal of Machine Learning Research}, vol.~23, no. 303, pp. 1--28, 2022.

\bibitem{kucar2004hypergraph}
D.~Kucar, S.~Areibi, and A.~Vannelli, ``Hypergraph partitioning techniques,'' \emph{Dynamics of Continuous Discrete and Impulsive Systems Series A}, vol.~11, pp. 339--368, 2004.

\bibitem{sorensen1999quantum}
\BIBentryALTinterwordspacing
A.~S{\o}rensen and K.~M{\o}lmer, ``Quantum computation with ions in thermal motion,'' \emph{Physical review letters}, vol.~82, no.~9, p. 1971, 1999. [Online]. Available: \url{https://journals.aps.org/prl/abstract/10.1103/PhysRevLett.82.1971}
\BIBentrySTDinterwordspacing

\bibitem{IonQ_blog}
``IonQ, Algorithmic Qubits: A better single number metric,'' \url{https://ionq.com/resources/ algorithmic-qubits-a-better-single-number-metric}.

\bibitem{IonQ_Aria}
``IonQ Aria,'' \url{https://ionq.com/quantum-systems/aria}.

\bibitem{IonQ_Forte}
``IonQ Forte,'' \url{https://ionq.com/quantum-systems/forte}.

\bibitem{lubinski2023application}
\BIBentryALTinterwordspacing
T.~Lubinski, S.~Johri, P.~Varosy, J.~Coleman, L.~Zhao, J.~Necaise, C.~H. Baldwin, K.~Mayer, and T.~Proctor, ``Application-oriented performance benchmarks for quantum computing,'' \emph{IEEE Transactions on Quantum Engineering}, 2023. [Online]. Available: \url{https://ieeexplore.ieee.org/abstract/document/10061574}
\BIBentrySTDinterwordspacing

\bibitem{maksymov2023enhancing}
\BIBentryALTinterwordspacing
A.~Maksymov, J.~Nguyen, Y.~Nam, and I.~Markov, ``Enhancing quantum computer performance via symmetrization,'' \emph{arXiv preprint arXiv:2301.07233}, 2023. [Online]. Available: \url{https://arxiv.org/abs/2301.07233}
\BIBentrySTDinterwordspacing

\bibitem{grimes1994shifted}
R.~Grimes, J.~Lewis, \& H. Simon, ``A shifted block Lanczos algorithm for solving sparse symmetric generalized eigenproblems``, \emph{SIAM Journal On Matrix Analysis And Applications}, vol.~15, pp.~228--272 (1994)

\bibitem{montanaro2016quantum}
A. Montanaro, ``Quantum algorithms: an overview,'' \emph{npj Quantum Information}, vol. 2, no. 1, pp. 1--8, 2016.

\bibitem{Guerreschi2019}
G. G. Guerreschi and M. Smelyanskiy, ``Practical optimization for hybrid quantum-classical algorithms,'' \emph{Quantum Science and Technology}, vol. 4, no. 1, p. 014001, 2019.

\bibitem{Preskill2018}
J. Preskill, ``Quantum computing in the NISQ era and beyond,'' \emph{Quantum}, vol. 2, p. 79, 2018.

\bibitem{Babbush2021}
R. Babbush, S. Boixo, J. Preskill, and H. Neven, ``Focus beyond quantum supremacy,'' \emph{Nature Physics}, vol. 16, pp. 105--109, 2020.



\end{thebibliography}

\section{Appendix A}

\subsection{Modified Fiduccia-Mattheyses Algorithm Integrated with VarQITE}
\label{SI:FM+VQITE}

In this section, we describe the details of our modified FM algorithm. Given a graph \(G(V,E)\), the original FM algorithm seeks partitions with minimal edge cuts through a local search guided by the Gain function \(D[v]\), efficiently computable as:
\begin{equation}
D[v] = \sum_{u \in N(v)} w(v, u) \cdot (-1)^{\text{same\_partition}(u, v)}
\end{equation}
Here, \(N(v)\) is the set of neighbors of vertex \(v\), \(w(v, u)\) denotes the edge weight connecting vertices \((u,v)\), and \(\text{same\_partition}(u,v)\) indicates whether two vertices belong to the same partition.

The original FM algorithm prioritizes node swaps based on the highest gain \(D[v]\), ensuring each swap either preserves or improves partition balance \cite{fiduccia1988linear}. This local search continues until no further improvement or until reaching a set maximum iteration count \(M\).

We propose a modification to enforce the nodal-weight balance constraint explicitly by prioritizing node removal from the partition with greater total nodal weight \(P[b]\), defined as:

\begin{equation}
P[b] =
\begin{cases}
P_0, & b \geq 0, \\
P_1, & b < 0.
\end{cases}
\end{equation}

This modification ensures each swap step reduces both edge cut and nodal-weight imbalance. By using VarQITE-generated partitions as initialization, the modified FM algorithm achieves significantly better convergence and solution quality.

Algorithm~\ref{FM_algorithm} presents the detailed steps of this modified FM procedure. The local search complements the global optimization from VarQITE, efficiently refining partitions and accelerating convergence toward near-optimal solutions, thus enhancing the quantum-classical hybrid optimization process.

\begin{algorithm}[hbt]
    \caption{Modified Fiduccia-Mattheyses Algorithm}
    \label{FM_algorithm}
    \begin{algorithmic}[1]
        \Require Graph $G(V, E)$, max iterations $M$, nodal weight imbalance tolerance $\epsilon$
        \Ensure Partition bitstring $\mathbf{p}$
        
        \State Initialize random bitstring $\mathbf{p}$, with length $n \gets |V|$ and equal cardinality
        \State Compute initial nodal weight balance $b$
                
        \For{$iteration$ = 1$ \textbf{ to } M$}
            \State Compute $D[v]$ for each node $v$ 
            \State Mark all nodes as "untouched"
            \State Set edge-cut gain $g \gets 0$
            
            \For{$step$ = 1$ \textbf{ to } n$}                
                \State $v^* \gets \arg\max_{v \in P[b], v \notin \text{touched}} D[v]$
                
                \If{no valid $v^*$ found}
                    \State \textbf{break}
                \EndIf
                
                \State Swap $v^*$ to the opposite partition
                \State Mark $v^*$ as ``touched''
                \State Update edge-cut gain $g \gets g + D[v^*]$
                \State Update weight balance $b$
                
                \State Update $D[v]$ for affected nodes
            \EndFor
            \State $g^* \gets $ largest $g$ in the current $iteration$ with $b\leq \epsilon$
            \If{$g^*\leq 0$}
                \State \textbf{break}
            \EndIf
            
            \State Revert $\mathbf{p}$ to the configuration that achieved $g^*$
        \EndFor
        
        \State \Return Final partition bitstring $\mathbf{p}$
    \end{algorithmic}
\end{algorithm}

\begin{figure}[hbt]
\includegraphics[width=1.0\linewidth]{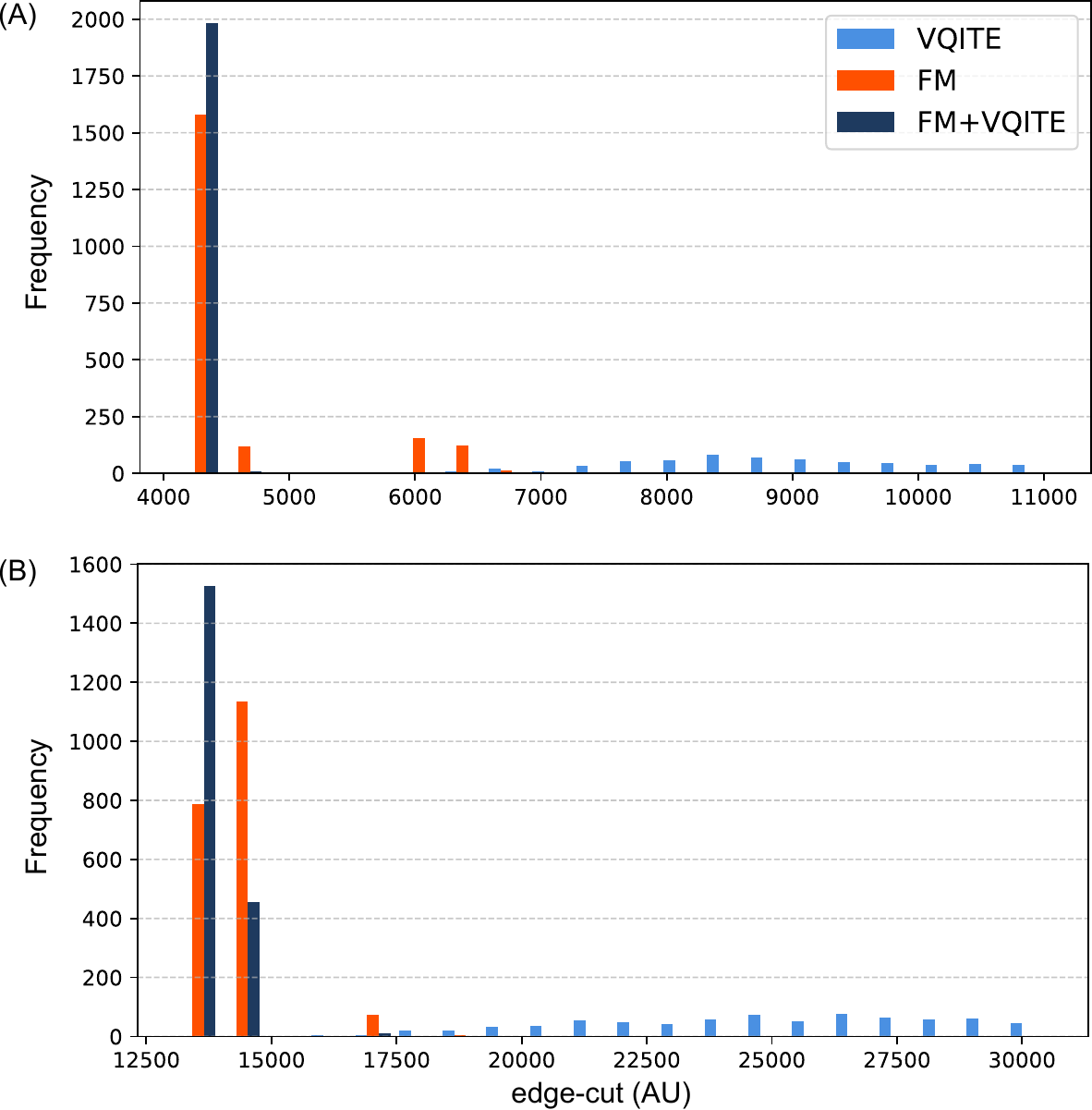}
\caption{Histograms of partition distributions binned by evaluated edge-cut cost for two benchmark problems: (A) RoofCrush and (B) BloodPump. The partitions are obtained using three different methods: FM (Fiduccia-Mattheyses) with random (equal-cardinality) initialization, VarQITE, and FM initialized with VarQITE partitions (FM+VarQITE). Each method generates 2,000 partitions, with VarQITE sampling 2,000 times, FM using 2,000 random initializations, and FM+VarQITE refining the 2,000 partitions from VarQITE. Partitions with $>5\%$ imbalance are filtered out per problem constraints. A method is considered superior if it yields more partitions in total and more partitions toward the left, indicating lower edge-cut costs.}
\label{fig:hist_FM+VQITE}
\end{figure}

Here we show the results from applying the modified FM algorithm to the VarQITE results from IonQ Forte. Fig~\ref{fig:hist_FM+VQITE} shows a comparison of the histograms from the raw VarQITE results and after refinining using the modified FM algorithm. The results also include using the modified FM algorithm alone from a random distribution of solutions of equal cardinality. The figure shows that a superior set of solutions can be found from the ``VarQITE+FM'' method as compared to using VarQITE or FM alone in terms of the frequency of the solutions of a specific edge-cut value achieved.

\subsection{Comparison of merit factors between VarQITE noiseless simulations and LS-DYNA - Level 2 nested dissection}
\label{SI:merit_factors_level2}

In this section, we show the final probability distributions after convergence of the VarQITE algorithm both on a noiseless simulator and on IonQ quantum hardware. We also show the merit factors computed from the VarQITE noiseless simulations of graphs extracted from a Level 2 nested dissection using LS-DYNA.

\begin{figure}[hbt]
\center \includegraphics[width=0.7\linewidth] {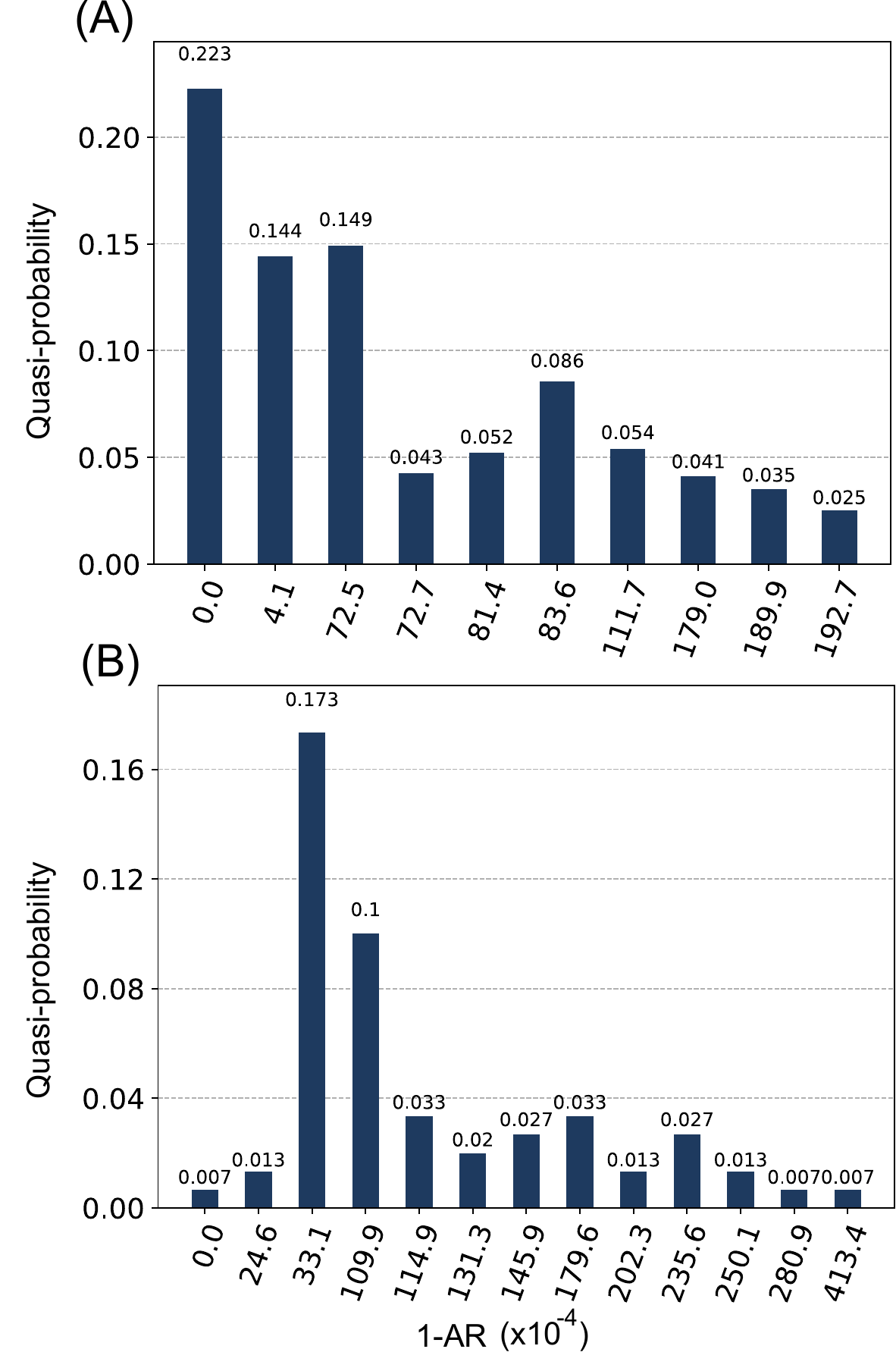}
    \caption{Probability histogram of a few unique solutions with the lowest objective function values from RoofCrush GPP problem for the graph with (A) 32 nodes (simulation) and (B) 30 nodes (IonQ Forte). The x-axis shows solutions with increasing values of 1-Approximation Ratio (AR). The optimal solution to the GPP problem is the first bar from the left, where the 1-AR value is 0.0.}
    \label{fig:RC_histogram}
\end{figure}

Figure~\ref{fig:RC_histogram} shows the sampled distributions from the last iteration of the VarQITE algorithm for RoofCrush GPP problem executed on noiseless simulator and on IonQ Forte. The first bar from the left represents the optimal solution to the problem with a 1-AR value of 0.0. Various other near-optimal solutions are also sampled.

\begin{figure}[hbt]
    \centering
    \includegraphics[width=\linewidth]{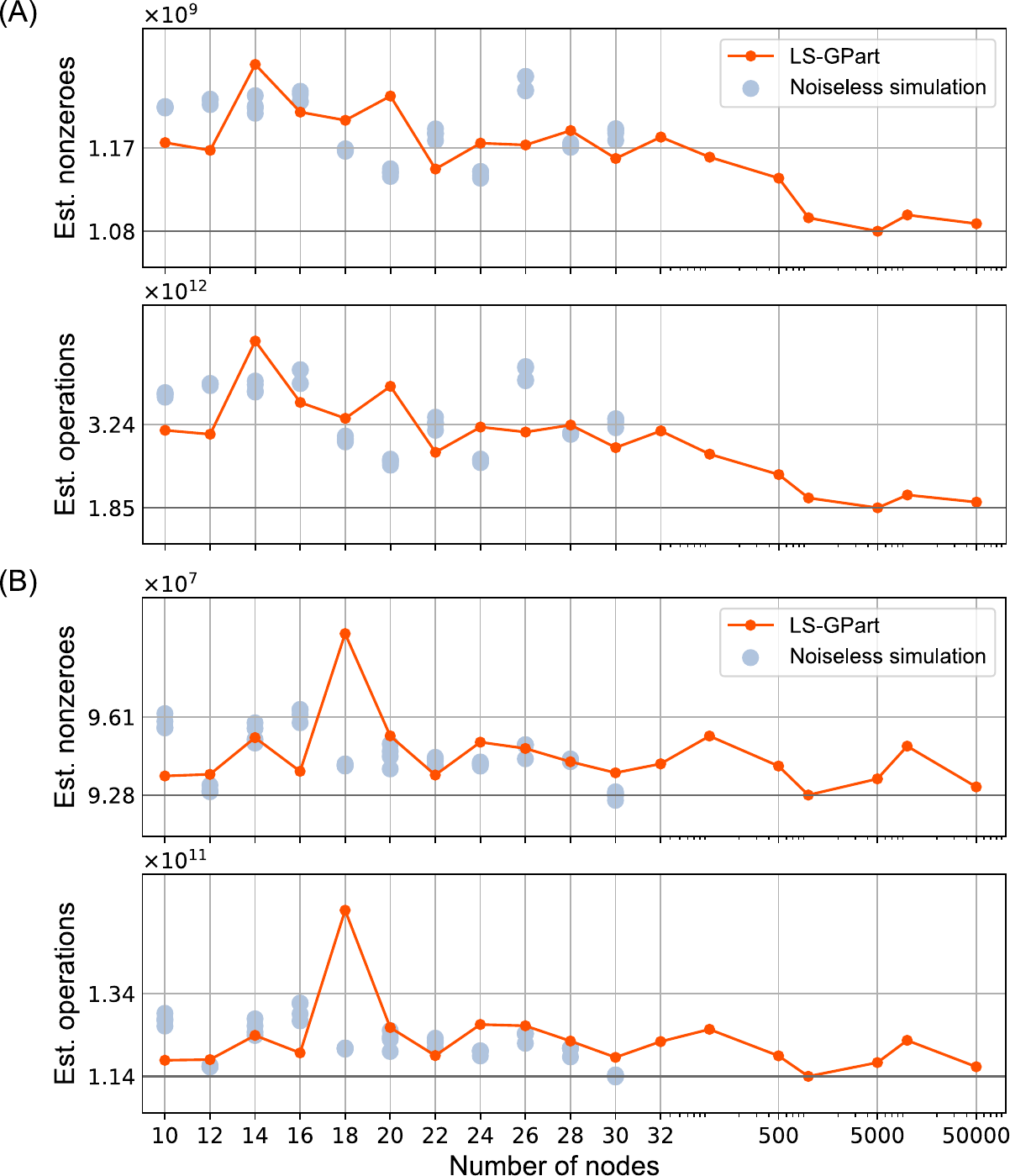}
    \caption{Plot of merit factors obtained from the VarQITE algorithm executed on (A) the RoofCrush problem and (B) the BloodPump problem at Level 2 nested dissection compared with the classical heuristic in LS-DYNA (LS-GPart): (top) The number of non-zeros (``fill-in''), (bottom) The number of operations estimated from symbolic factorization by LS-DYNA. The VarQITE data is only from graph partitions which maintain a nodal weight balancewithin a 5\% tolerance. The QPU data is from graphs with the same number of nodes as the corresponding simulation data, just offset from the simulation points for clarity. The plot also shows the results of applying the modified FM algorithm to the QPU data. This plot should be compared to Fig.~\ref{fig:merit_factors-L1} in that it shows similar behavior for the BloodPump case, where solutions coarsened to 30 nodes/qubits are competitive with LS-GPart at up to 10,000 nodes.}
    \label{fig:merit_factors-L2}
\end{figure}

Figure~\ref{fig:merit_factors-L2} shows the comparison of merit factors from noiseless simulations of the VarQITE algorithm on Level 2 nested dissection graphs and LS-GPart for the RoofCrush and BloodPump problems. The figure shows encouraging results as there are simulation points that show lower merit factors than LS-GPart, especially for the BloodPump problem. There is an additional nuance here since the results depend on the specific solution chosen for the Level 1 graph partitioning. There are many cases where LS-GPart chooses a Level 1 solution that is outside the 5\% nodal weight balanace which can result in a lower merit factor for Level 2. But the VarQITE solutions always respect the 5\% nodal weight balance. This makes the data shown in Fig~\ref{fig:merit_factors-L2} a little more noisy for the RoofCrush problem. 

\subsection{Additional data: measuring wall clock time for linear solve}
\label{SI:WCT}

\begin{figure}[hbt]
    \centering
    \includegraphics[width=\linewidth]{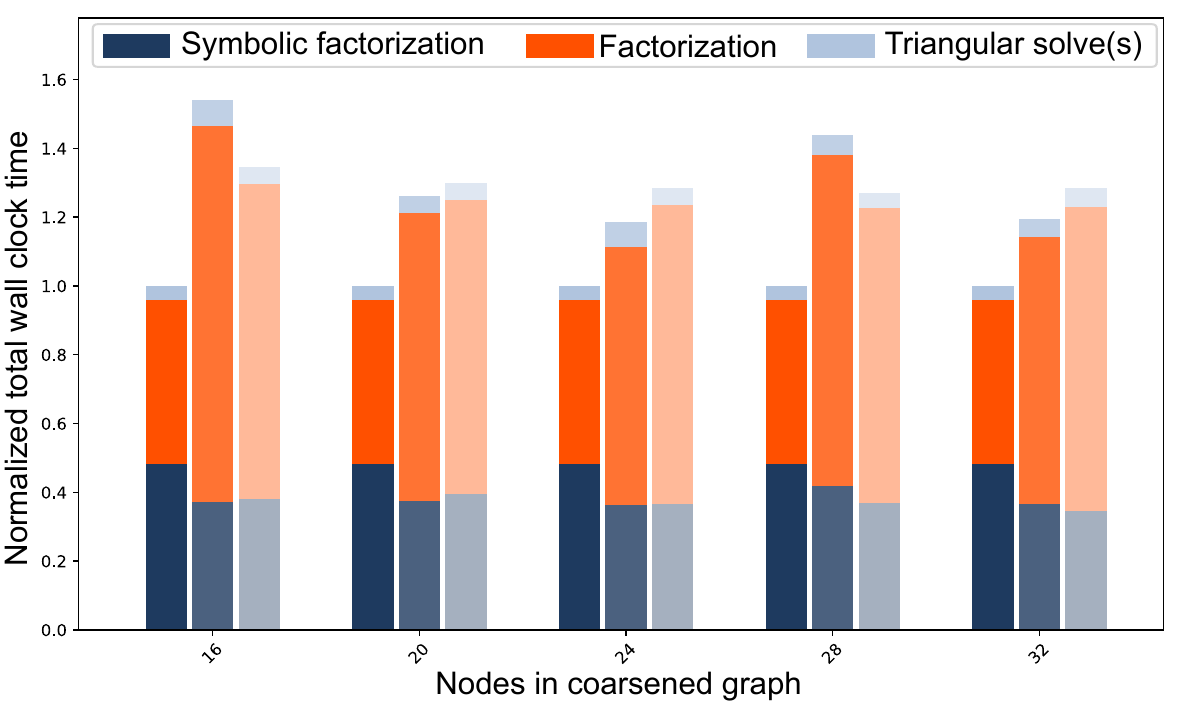}
    \caption{Total WCT comparison for linear system solve of the RoofCrush problem on 16 MPI ranks. The leftmost bar represents total WCT when coarsening the graph to 10,000 nodes (production setting). The middle and right bars compare WCT when coarsening the graph to different numbers of nodes shown on the x-axis, and using the internal vs. the external partitioner}
    \label{fig:RoofCrush_WCT}
\end{figure}

\begin{figure}[hbt]
    \centering
    \includegraphics[width=\linewidth]{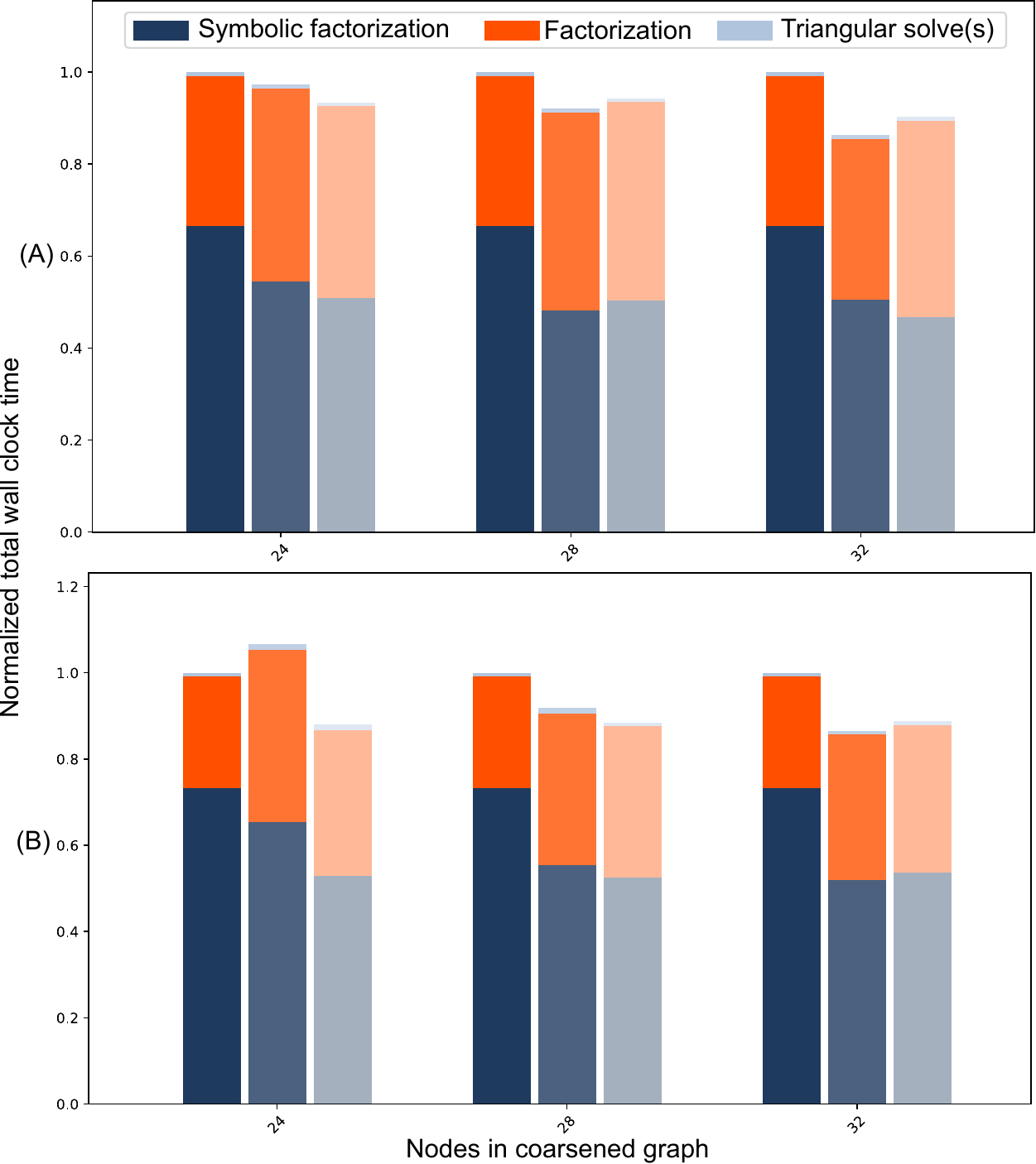}
    \caption{Total WCT comparison for linear system solve of the BloodPump problem on A) 8 MPI ranks and B) 16 MPI ranks. In each group, the leftmost bar represents total WCT when coarsening the graph to 10,000 nodes (production setting). The middle and right bars compare WCT when coarsening the graph to different numbers of nodes shown on the x-axis, and using the internal vs. the external partitioner}
    \label{fig:BloodPump_WCT}
\end{figure}

\begin{figure}[hbt]
    \centering
    \includegraphics[width=\linewidth]{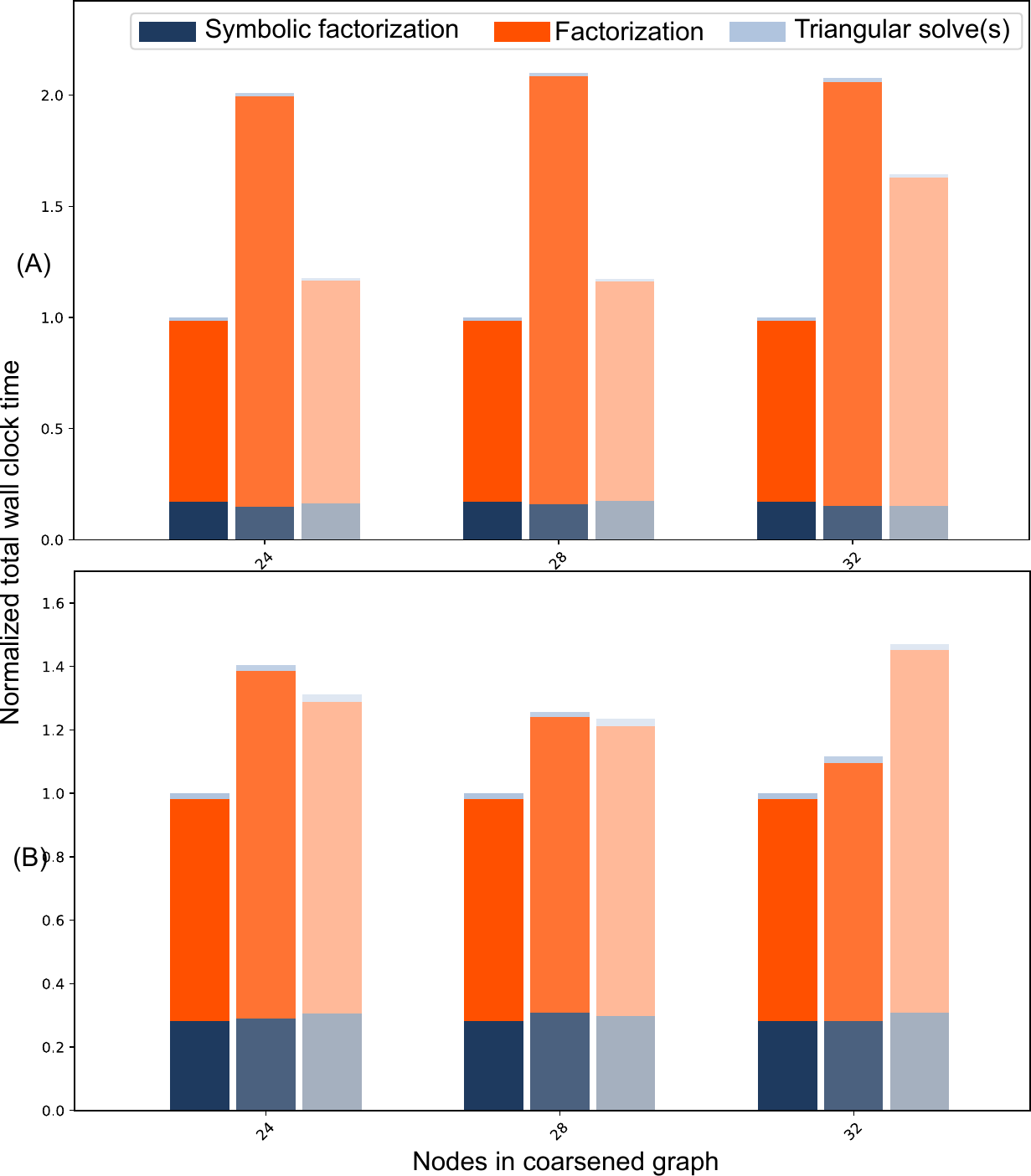}
    \caption{Total WCT comparison for linear system solve of the VibrationAnalysis problem on A) 8 MPI ranks and B) 24 MPI ranks. In each group, the leftmost bar represents total WCT when coarsening the graph to 10,000 nodes (production setting). The middle and right bars compare WCT when coarsening the graph to different numbers of nodes shown on the x-axis, and using the internal vs. the external partitioner}
    \label{fig:HondaAccord_WCT}
\end{figure}

Here we show additional wall clock time experiments from the RoofCrush, BloodPump and VibrationalAnalysis problems. Figures~\ref{fig:RoofCrush_WCT}, ~\ref{fig:BloodPump_WCT} and ~\ref{fig:HondaAccord_WCT} show the results from running on different numbers of MPI ranks. The results still show a largely decreasing trend of WCT from VarQITE solutions with increasing number of nodes of the coarsened graph. The VarQITE solutions also show an improvement over LS-GPart in most cases, but not all. The cases where VarQITE does not show an improvement over LS-GPart need further investigation with likely causes ranging from incomplete convergence of the VarQITE algorithm to control over the compute environment while computing WCT.

\end{document}